\documentclass[twocolumn]{autart}    
\usepackage{float}
\usepackage{amsmath}
\usepackage{amssymb} 

\usepackage{subfigure}

\usepackage{graphicx}

\usepackage{bm}

\usepackage{array}

\usepackage{chngcntr}

\newtheorem{remark}{Remark}
\newtheorem{proposition}{Proposition}

\newtheorem{theorem}{Theorem}
\counterwithin*{theorem}{section} 

\usepackage{cite,notoccite}									
							
\usepackage[dvipsnames]{xcolor}	

\begin{document}

\begin{frontmatter}

\title{Extremum Seeking Control with Attenuated Steady-State Oscillations\thanksref{footnoteinfo}}

\thanks[footnoteinfo]{This research was supported by the Office of Naval Research under Grant No. N00014-18-1-2215. This paper was not presented at any IFAC meeting. Corresponding author Kamesh Subbarao. Tel. +1-817-272-7467. 
Fax +1-817-272-5010.}

\author{Diganta Bhattacharjee}\ead{diganta.bhattacharjee@mavs.uta.edu},    
\author{Kamesh Subbarao}\ead{subbarao@uta.edu}               

\address{Aerospace Systems Laboratory, Department of Mechanical and Aerospace Engineering, The University of Texas at Arlington, Arlington, TX, 76019, USA}


\begin{abstract}                          
We propose two perturbation-based extremum seeking control (ESC) schemes for general single input single output nonlinear dynamical systems, having structures similar to that of the classical ESC scheme. We propose novel adaptation laws for the excitation signal amplitudes in each scheme that drive the amplitudes to zero. The rates of decay for both the laws are governed by the gradient measures of the unknown reference-to-output equilibrium map. We show that the proposed ESC schemes achieve practical asymptotic convergence to the extremum with a proper tuning of the parameters in the proposed schemes. As the extremum is reached, and the magnitudes of the gradient measures become small, the excitation signal amplitudes converge to zero. Thus, the proposed schemes ensure that the excitation signal is attenuated as the system output converges to a neighborhood of the extremum and the steady-state oscillations about the extremum, typically observed for the classical ESC schemes, are attenuated. Simulation examples are included to illustrate the effectiveness of the proposed schemes. 
\end{abstract} 

\end{frontmatter}


\section{Introduction}
Extremum seeking control (ESC) is a form of adaptive control technique that optimizes the steady-state performance of a dynamical system (Haring and Johansen \cite{Haring_and_Johansen_2017}, \cite{Haring_and_Johansen_2018}) and drives the steady-state system output to its extremum by systematically tuning the reference input to the system (Krstic and Wang \cite{Krstic_Wang_2000}, Tan et al. \cite{Tan_et_al_2006}). The objective of ESC is not only to extremize the output but also to keep the output at the extremum value. The control scheme does not rely upon the explicit knowledge of the reference-to-output map and relies instead on the measured output value. Thus, ESC is a model-free optimization technique that ensures optimal steady-state operation, given there exists an extremum in the reference-to-output map. In particular, the classical ESC schemes, proposed by Krstic and Wang \cite{Krstic_Wang_2000} and Tan et al. \cite{Tan_et_al_2006}, deal with finding the extremum of the reference-to-output \textit{equilibrium} map where the relationship between the reference and the steady-state output is characterized by a nonlinear, static mapping. 

Although the extremum seeking problem had been studied since 1950s and 1960s, the first rigorous stability proof of an ESC scheme was provided by Krstic and Wang \cite{Krstic_Wang_2000} in 2000. That has led to a renewed interest in the theoretical developments and applications of ESC in a variety of disciplines. Especially, due to the attractive feature of model-free optimization, ESC has been applied to engineering problems involving a highly complex system that is hard to model accurately or to problems for which the input-to-output mapping is completely or partially unknown. Some applications of ESC to such problems include antilock braking systems (Drakunov et al.\cite{Drakunov_et_al}, Nesic et al. \cite{Nesic_et_al_2012}, Wang et al. \cite{Wang_et_al}), power reduction, induced drag minimization, or lift maximization of the wingman aircraft in formation flight (Binetti et al. \cite{Binetti_et_al}, Chichka et al. \cite{Chichka_et_al}, Brodecki and Subbarao \cite{Brodecki_Subbarao}), stirred-tank reactors (Dehaan and Guay \cite{Dehaan_Guay_2005}, Adetola and Guay \cite{Adetola_Guay_2007}), electromagnetic actuators (Benosman and Atinc \cite{Benosman_Atinc_2015}), real-time optimization over a network of dynamic agents (Poveda and Quijano \cite{Poveda_Quijano_2015}, Guay et al. \cite{Guay_et_al_2018}), and source localization using unmanned aerial vehicles and mobile robots (Bhattacharjee et al. \cite{Bhattacharjee_et_al}, Ghods and Krstic \cite{Ghods_Krstic}). It is also worth mentioning that recently a different class of extremum seeking controllers has emerged in the literature that utilizes the so-called Lie bracket system (Durr et al. \cite{Durr_et_al_2013, Durr_et_al_2017}, Grushkovskaya et al. \cite{Grushkovskaya_et_al_2018}, Suttner \cite{Suttner_2019}).  


\textbf{Motivation:} The essential idea behind ESC is to somehow extract the gradient information of the unknown reference-to-output map by introducing a periodic perturbation in the reference, termed excitation signal or dither signal. To extract the gradient information, a filtering-based approach is taken in the classical ESC schemes (Ariyur and Krstic \cite{Ariyur_Krstic_2003}, Krstic and Wang \cite{Krstic_Wang_2000}, Tan et al. \cite{Tan_et_al_2006, Tan_et_al_2010}). Alternatively, there are ESC schemes in the existing literature that involve estimating the gradient of this unknown function using an estimator or observer (Chichka et al. \cite{Chichka_et_al}, Haring and Johansen \cite{Haring_and_Johansen_2017}). The gradient information is then utilized to drive the reference so that the steady-state output is extremized. In the presence of multiple extrema, this approach might make the output converge to a local extremum instead of the global extremum. Tan et al. \cite{Tan_et_al_2009} proposed a solution to this issue by introducing an adaptation law for the excitation signal amplitude. They proposed an adaptation law that introduced a slowly decaying excitation signal with a large enough initial amplitude. However, the adaptation in the amplitude is solely dependent on the amplitude itself and not on the on-line optimization process, integral to the ESC method. Additionally, the scheme in Tan et al. \cite{Tan_et_al_2009} exhibits slow convergence and steady-state oscillations. In fact, this is a feature of all the classical ESC schemes where the system output and the reference exhibit steady-state oscillations about their respective extremum values.
One possible remedy to this issue of steady-state oscillations in classical ESC schemes was given in Wang et al. \cite{Wang_et_al} where they proposed a classical ESC based scheme that showed fast convergence to the extremum without any steady-state oscillations. Although the nominal part of the reference in Wang et al. \cite{Wang_et_al} is driven by a measure of the gradient (the demodulated signal), the amplitude of the excitation signal is driven to zero based on the error in the extremum estimation. We are interested in attenuating the steady-state oscillations for the classical ESC framework proposed by Krstic and Wang \cite{Krstic_Wang_2000}. This can be achieved by driving the excitation signal amplitude to zero as the extremum is reached. To this end, decay in the amplitude can be governed by a measure of the derivatives or gradients of the function to be optimized or extremized. This idea has been utilized for the ESC scheme in Moase et al. \cite{Moase_et_al_2010} and serves as a motivation for the proposed schemes.


\textbf{Contribution:} In this paper, we propose extremum seeking schemes motivated by the classical ESC methods that would drive the excitation signal amplitude to zero based on \textit{a measure of the gradient}. To that end, we take an approach that combines the classical schemes and the estimator or observer-based schemes. In doing so, we propose two extremum seeking schemes  having structures similar to that of the classical scheme proposed by Krstic and Wang \cite{Krstic_Wang_2000}, but with distinct adaptation laws. These two adaptation laws for the excitation signal amplitude are motivated by the approach introduced by Tan et al. \cite{Tan_et_al_2009}. However, in contrast to Tan et al. \cite{Tan_et_al_2009}, the rate of decay of the amplitude is governed by the gradient measures of the unknown reference-to-output equilibrium map. One of the proposed adaptation laws utilizes an estimated value of the gradient wherein the gradient estimates are obtained using a Kalman Filter, similar to the procedure described by Chichka et al. \cite{Chichka_et_al}. For the other law, we utilize an equivalent measure of the gradient. It is argued in Krstic and Wang \cite[Section 3]{Krstic_Wang_2000} that the low-pass filtered demodulated signal ($\xi$ in Fig. 1 and 2 in Krstic and Wang \cite{Krstic_Wang_2000}) represents a variable equivalent to the scaled gradient of the reference-to-output equilibrium map. That signal is utilized in the definition of the second adaptation law. We show that the proposed schemes achieve practical asymptotic convergence to the extremum.  


\textbf{Organization:} The assumptions and problem formulation are described in Section \ref{section-2}. Section \ref{section-3} elaborates the main results and illustrative simulation examples are given in Section \ref{section-4}. Finally, Section \ref{section-5} summarizes the findings of this paper and provides concluding remarks.

\section{Preliminaries and problem formulation} \label{section-2}
To be consistent with the existing results in the literature, we adopt the same notations and a similar problem formulation as given in Krstic and Wang \cite{Krstic_Wang_2000} and Tan et al. \cite{Tan_et_al_2006, Tan_et_al_2009}. For the sake of completeness, we describe the problem formulation and the list of assumptions in this section. Consider a general single input and single output (SISO) nonlinear dynamical model given by
\begin{equation}
\dot{\bm{x}} = \bm{f}(\bm{x},u), \hspace{0.5cm} y = h(\bm{x}),
\end{equation}
where $\bm{x} \in \mathbb{R}^n$ is the state vector, $u \in \mathbb{R}$ is the control input, $y \in \mathbb{R}$ is the measured output, and $\bm{f} : \mathbb{R}^n \times \mathbb{R} \to \mathbb{R}^n $ and $h : \mathbb{R}^n \to \mathbb{R} $ are smooth. 

\begin{figure}[h]
\begin{center}
\includegraphics[width = 0.45\textwidth]{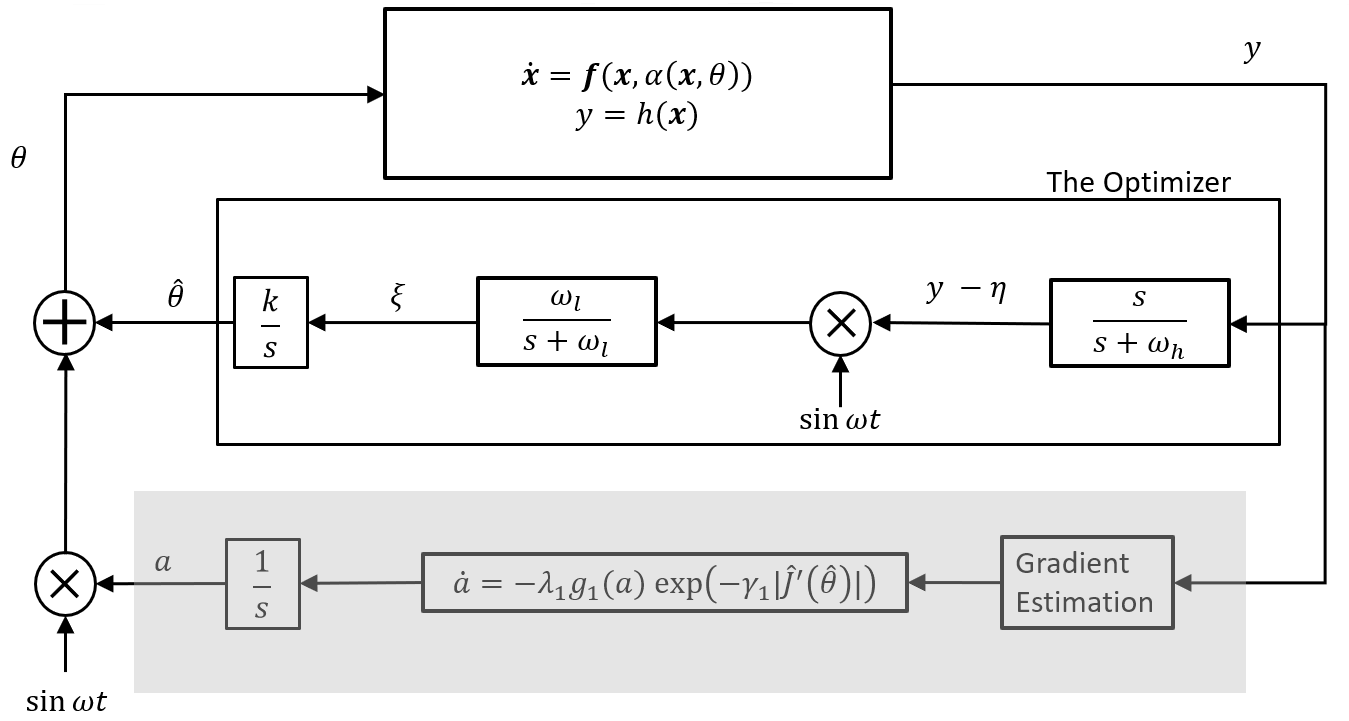}      
\caption{Proposed ESC scheme-1}            			   
\label{fig1}                                   
\end{center}                                   
\end{figure}

\begin{figure}[h]
\begin{center}
\includegraphics[width = 0.45\textwidth]{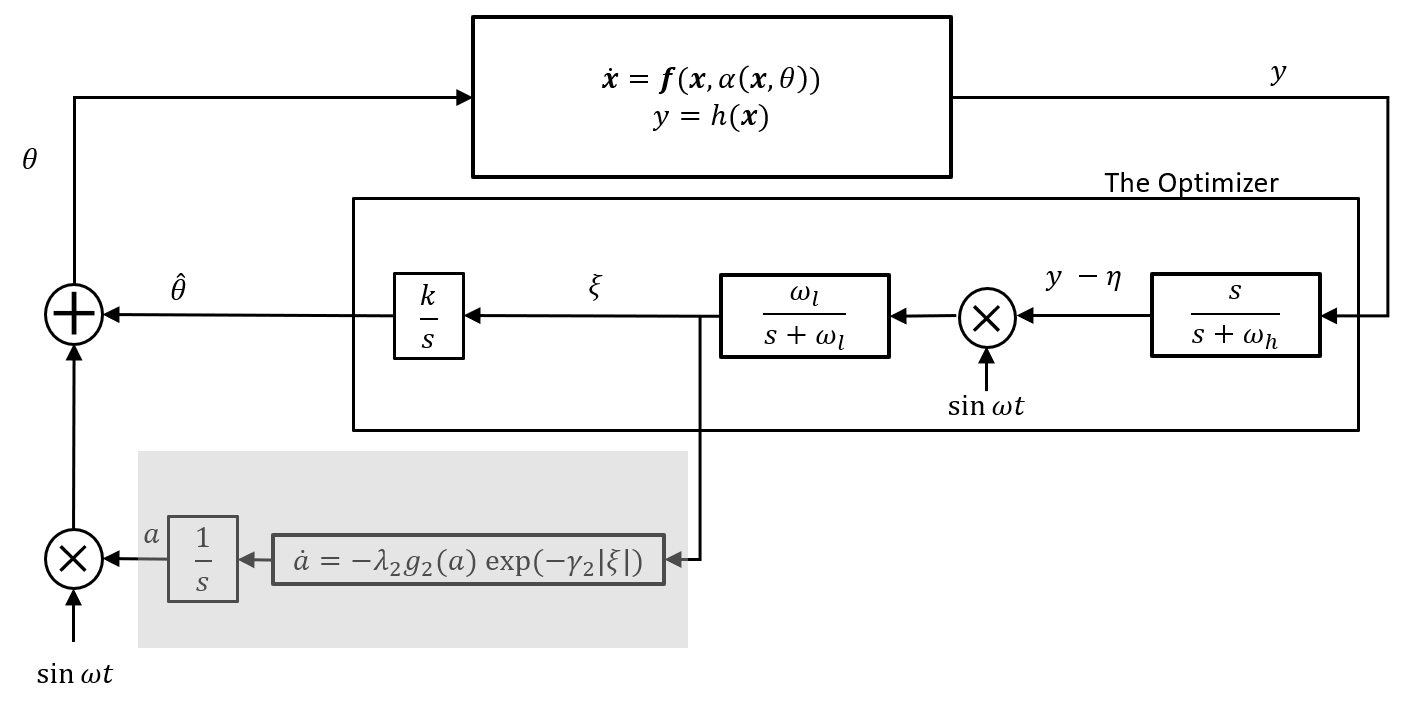}    
\caption{Proposed ESC scheme-2}                         
\label{fig2}                                   
\end{center}                                   
\end{figure}

Suppose there exists a family of smooth feedback control laws of the form 
\begin{equation}
u = \alpha(\bm{x},\theta),
\end{equation} 
parameterized by the scalar parameter $\theta$. Then, the closed-loop system 
\begin{equation} \label{plant dynamics}
\dot{\bm{x}} = \bm{f} (\bm{x}, \alpha(\bm{x},\theta)) 
\end{equation}
has equilibria parameterized by $\theta$. In this paper, we are interested in investigating the extremum seeking schemes shown in Figs. \ref{fig1} and \ref{fig2}. Note that one can retrieve the classical ESC scheme given in Krstic and Wang \cite{Krstic_Wang_2000} by (a) removing the input to the shaded blocks, (b) replacing the shaded blocks in the proposed schemes with a constant amplitude $a$ for the excitation signal, and (c) multiplying the demodulation signal with the same amplitude $a$. Moreover, $\hat{\theta}$ in these schemes can be considered to be the nominal part of the reference or the current estimate of the extremum $\theta^{\star}$ and $a \sin \omega t$ to be the excitation signal with the amplitude $a = a(t)$ as a function of time (Krstic and Wang\cite{Krstic_Wang_2000}, Haring and Johansen \cite{Haring_and_Johansen_2017}, \cite{Haring_and_Johansen_2018}).

We make same assumptions for the closed-loop system as Krstic and Wang \cite{Krstic_Wang_2000}. These are given as follows.
\begin{assum} \label{Assumption 1}
There exists a smooth function $\bm{l} : \mathbb{R} \to \mathbb{R}^n$ such that $\bm{f} (\bm{x}, \alpha(\bm{x},\theta)) = 0$, if and only if $\bm{x} = \bm{l}(\theta)$.
\end{assum} 
\begin{assum}  \label{Assumption 2}
For each $\theta \in \mathbb{R}$, the equilibrium $\bm{x} = \bm{l}(\theta)$ of system \eqref{plant dynamics} is locally exponentially stable, uniformly in $\theta$.
\end{assum}
This assumption means that we have a control law that would stabilize the system locally, irrespective of the modeling knowledge of either $\bm{f}(\bm{x},u)$ or $\bm{l}(\theta)$. Without loss of generality, we consider the problem of maximizing the steady-state output by finding the maximum in the output equilibrium map $y = h(\bm{l}(\theta))$. The case for the minimization problem can be treated similarly by replacing $y$ with $-y$. Since we are interested in finding the maximum in the output equilibrium map $y = h(\bm{l}(\theta))$, let us denote $J (\theta) = h(\bm{l}(\theta))= (h \circ \bm{l}) (\theta)$ as the objective function for the extremum seeking problem. Similarly, for the minimization problem, $-h(\bm{l}(\theta))$ can be treated as a cost function that needs to be minimized.
\begin{assum} \label{Assumption 3}
There exists $\theta^{\star} \in \mathbb{R}$ such that 
\begin{equation}
\begin{split}
J^{\prime} (\theta^{\star}) &= 0, \\
J^{\prime \prime} (\theta^{\star}) &< 0. 
\end{split}
\end{equation}
\end{assum}
This last assumption implies that the objective function $J(\theta)$ has a maximum at $\theta = \theta^{\star}$. As shown in Fig. \ref{fig1}, the gradient of the objective function has to be estimated. For that, we have adapted the Kalman Filter based gradient estimation scheme from Chichka et al. \cite{Chichka_et_al} and we denote the gradient estimate at $\hat{\theta}$ as $\hat{J}^{\prime}(\hat{\theta})$. The Kalman Filter `truth' model for this case is as following.
\begin{equation} \label{Kalman Filter}
\dot{\bm{\psi}} = \begin{bmatrix}
0 & \omega & 0 \\
-\omega & 0 & 0 \\
0 & 0 & 0 
\end{bmatrix} \bm{\psi} + \bm{w},\ 
\tilde{y} = \begin{bmatrix}
a & 0 & 1
\end{bmatrix} \bm{\psi} + v, 
\end{equation}
where $\bm{\psi} = \hspace{-0.1cm} \left(\psi_1, \psi_2, \psi_3 \right) \hspace{-0.1cm} = \hspace{-0.1cm} ( J^{\prime}(\hat{\theta}) \sin \omega t, J^{\prime}(\hat{\theta}) \cos \omega t,  J (\hat{\theta}) )$, $\omega$ is the excitation signal frequency, $a$ is the excitation signal amplitude, $\bm{w} \sim \mathcal{N}(\bm{0},\bm{Q})$ and $v \sim \mathcal{N}(0,r)$ are zero-mean Gaussian white-noise terms with covariances $\bm{Q}$ and $r$, respectively, and $\tilde{y}$ is a measurement of the objective function at $\theta = \hat{\theta} + a \sin \omega t$. Furthermore, $\bm{w}$ and $v$ are uncorrelated. The estimate of $\bm{\psi}$ is denoted by $\hat{\bm{\psi}} = (\hat{\psi}_1, \hat{\psi}_2, \hat{\psi}_3)$. Therefore, the estimated gradient magnitude is given by $|\hat{J}^{\prime}(\hat{\theta})| = \sqrt{\hat{\psi}_1^2 + \hat{\psi}_2^2}$. Note that we do not specify a sign to the gradient estimate and only utilize the estimated gradient magnitude (cf. \eqref{adaptation law-1}). 

\begin{assum} \label{Assumption 4}
There exists a positive constant $\epsilon_0$ such that the estimates satisfy $|\hat{J}^{\prime}(\hat{\theta})| \leq \epsilon_0$ for all $t \geq t_0 \geq 0$. Moreover, there exists a small positive constant $\epsilon_J$ and a time interval $\Delta T > 0$ such that $| \ |J^{\prime}(\hat{\theta})| - |\hat{J}^{\prime}(\hat{\theta})| \ | \leq \epsilon_J$ for all $t \geq t_0 + \Delta T$. 
\end{assum} 
Assumption \ref{Assumption 4} implies that the Kalman Filter is performing adequately after some non-zero time interval $\Delta T$ from time $t_0$, when the scheme was initialized. 

\section{Main results}   \label{section-3}
In this paper, we propose two adaptation laws for the amplitude of the excitation signal so that the excitation signal converges to zero as the extremum is reached. These are given as following.
\begin{itemize}
\item Adaptation law (scheme-1): \begin{equation} \label{adaptation law-1}
\dot{a} = -\lambda_1 \ g_1 (a) \ \text{exp}(-\gamma_1 |\hat{J}^{\prime}(\hat{\theta})|), \hspace{0.2cm} a(t_0) = a_0 > 0,
\end{equation} 
\item Adaptation law (scheme-2): \begin{equation} \label{adaptation law-2}
\dot{a} = -\lambda_2 \ g_2 (a) \ \text{exp}(-\gamma_2 |\xi|), \hspace{0.2cm} a(t_0) = a_0 > 0,
\end{equation} 
\end{itemize}
where $\lambda_1 >0$ and $\lambda_2 >0$ are design parameters, $\gamma_1$ and $\gamma_2$ are $O(1)$ positive scaling parameters, $\xi$ is as shown in Fig. \ref{fig2}, and $g_1 (a), g_2 (a)$ are locally Lipschitz functions that are zero at zero and positive otherwise. The parameters $\gamma_1$, $\gamma_2$, and $a_0$ are to be selected based on the problem at hand and are left to the choice of the designer.
\begin{remark} \label{Remark 1} 
The choice of the adaptation laws in \eqref{adaptation law-1} and \eqref{adaptation law-2} is motivated by the adaptation law proposed in Tan et al. \cite{Tan_et_al_2009}. We remark that the proposed laws are equivalent to the one in Tan et al. \cite{Tan_et_al_2009} when the gradient measures are sufficiently small. However, since the standing assumption in ESC is that the extremum is unknown, one cannot guarantee that the gradient measures are small when an ESC scheme is initialized. To this end, the scaling parameters $\gamma_1$ and $\gamma_2$ for scheme-1 and scheme-2, respectively, can be chosen sufficiently large so that the rates of decay in the amplitudes are initially governed by the exponents in \eqref{adaptation law-1} and \eqref{adaptation law-2}. This means that the decay in the amplitudes is seized during the initialization period of the proposed schemes and is similar, in spirit, to the requirement of a sufficiently large $a_0$ for the scheme in Tan et al. \cite{Tan_et_al_2009}. In fact, for the proposed schemes, $a_0$ has to be sufficiently small (cf. Theorems 1 and 2).
\end{remark}

\begin{remark}  \label{Remark on the two phase decay in the excitation signal amplitude}
Utilzing the discussion in Remark \ref{Remark 1}, we choose $\gamma_1$ and $\gamma_2$ sufficiently large to arrest the rates of decay in the amplitudes during the initialization period of the proposed schemes and allow the optimizer to make proper corrections (see Section \ref{section-4}). Then, as the system approaches the extremum and the gradient measures become sufficiently small, the rates of decay are approximately governed by the adaptation law in Tan et al. \cite{Tan_et_al_2009}, i.e., $\dot{a} \approx - \lambda_i \ g_i (a) \ (i=1,2)$. Thus, for the proposed schemes, we would like the decay in the excitation signal amplitudes to occur in two phases, with $\gamma_1$ and $\gamma_2$ properly chosen (see Section \ref{section-4}). A switching control-based strategy for the decay in the excitation signal amplitude is given in Moura and Chang \cite{Moura_Chang_2013}. However, the switching requires a Lyapunov function that utilizes the knowledge of an accurate enough `nominal' extremum and the corresponding numerical values of the objective function derivatives as well as a switching threshold. Note that no such prior knowledge is required for the proposed schemes in our paper.
\end{remark} 

Next, we elaborate the stability analysis of the proposed ESC schemes shown in Figs. 1 and 2. Letting $\tilde{\theta} = \hat{\theta} - \theta^{\star}$, $\tilde{\eta} = \eta - J(\theta^{\star})$, and substituting $y = h(\bm{x})$, the closed-loop systems can be expressed as following.
\begingroup
\allowdisplaybreaks
\begin{align} \label{closed-loop system-1-2}
\text{Scheme-1:} \nonumber \\
\dot{\bm{x}} &= \bm{f} \left( \bm{x}, \alpha(\bm{x}, \tilde{\theta} + \theta^{\star} + a \sin \omega t) \right), \nonumber \\
\dot{\tilde{\theta}} &= k \xi, \nonumber \\
\dot{\xi} &= - \omega_l \xi + \omega_l (h(\bm{x}) - \tilde{\eta} - J(\theta^{\star})) \sin \omega t, \\
\dot{\tilde{\eta}} &= -\omega_h \tilde{\eta} + \omega_h (h(\bm{x}) - J(\theta^{\star})), \nonumber\\
\dot{a} &= -\lambda_1 \ g_1 (a) \ \text{exp}(-\gamma_1 |\hat{J}^{\prime}(\tilde{\theta} + \theta^{\star})|). \nonumber
\end{align}
\endgroup
\vspace{-0.75cm}
\begingroup
\allowdisplaybreaks
\begin{align} \label{closed-loop system-2-2}
\text{Scheme-2:} \nonumber \\
\dot{\bm{x}} &= \bm{f} \left( \bm{x}, \alpha(\bm{x}, \tilde{\theta} + \theta^{\star} + a \sin \omega t) \right), \nonumber \\
\dot{\tilde{\theta}} &= k \xi, \nonumber \\
\dot{\xi} &= - \omega_l \xi + \omega_l (h(\bm{x}) - \tilde{\eta} - J(\theta^{\star})) \sin \omega t, \\
\dot{\tilde{\eta}} &= -\omega_h \tilde{\eta} + \omega_h (h(\bm{x}) - J(\theta^{\star})), \nonumber\\
\dot{a} &= -\lambda_2 \ g_2 (a) \ \text{exp}(-\gamma_2 |\xi|). \nonumber
\end{align}
\endgroup
We introduce the following representation for the parameters in \eqref{closed-loop system-1-2} and \eqref{closed-loop system-2-2}. 
\begin{eqnarray*}
\omega_h &=& \omega \omega_H = \omega \delta \omega^{\prime}_H = O(\omega \delta), \\
\omega_l &=& \omega \omega_L = \omega \delta \omega^{\prime}_L = O(\omega \delta), \\
k &=& \omega K = \omega \delta K^{\prime}= O(\omega \delta), \\
\lambda_1 &=& \omega \lambda_{1_{1}} = \omega \delta \epsilon \lambda_1^{\prime} = O(\omega \delta \epsilon), \\
\lambda_2 &=& \omega \lambda_{2_{1}} = \omega \delta \epsilon \lambda_2^{\prime} = O(\omega \delta \epsilon),
\end{eqnarray*} 
where $\omega$, $\delta$, and $\epsilon$ are small positive constants and $\omega^{\prime}_H$, $\omega^{\prime}_L$, $K^{\prime}$, $\lambda_1^{\prime}$, and $\lambda_2^{\prime}$ are $O(1)$ positive constants. From the above representation of the parameters involved, we can conclude that the closed-loop systems of the proposed schemes should exhibit four time scales. These requirements on the time-scale properties of the proposed schemes are similar to the ones introduced in Krstic and Wang \cite{Krstic_Wang_2000} and Haring and Johansen \cite{Haring_and_Johansen_2017}, \cite{Haring_and_Johansen_2018}. These time scales are given by:
\begin{itemize}
\item fast - the system with the controller
\item medium fast - the periodic perturbations
\item medium slow - the filters in the proposed schemes
\item slow - the adaptation in the excitation signal amplitude
\end{itemize}
The system is required to be fast compared to the rest of the components of the schemes so that the difference between true output of the system and the steady-state output corresponding to the objective function remains small. The filters are required to be slower compared to the periodic perturbations as that would allow the filters to accurately estimate the nominal part of the reference ($\hat{\theta}$). Also, adaptation in the excitation signal amplitude is required to be sufficiently slow so that the optimality of the current estimate $\hat{\theta}$ is checked and appropriate corrections are made by the optimizer (see Remark \ref{Remark about the small corrections at the end}).
\begin{remark} \label{Remark about the small corrections at the end}
It follows from the analysis in Krstic and Wang \cite{Krstic_Wang_2000} that $\omega$ and $\delta$ should be sufficiently small for the proposed schemes. Moreover, $\epsilon$ has to be small, as in Tan et al. \cite{Tan_et_al_2009}. An analysis similar to the one recently given in Atta and Guay \cite{Atta_Guay_2019} can be carried out to establish the existence of equilibrium manifolds for the systems \eqref{closed-loop system-1-2} and \eqref{closed-loop system-2-2}. On these manifolds, we have $a=0$ and $\tilde{\theta} = \theta_c$ where $\theta_c$ is a constant, not necessarily zero. As the system approaches the extremum and the exponents in \eqref{adaptation law-1} and \eqref{adaptation law-2} are approximately equal to one, the rates of decay in the amplitudes are governed by $- \lambda_i \ g_i (a) \ (i=1,2)$. By making $\epsilon$ small and reducing the rates of decay of the amplitudes while the system approaches the extremum, we allow the optimizer to make corrections so that we have $\tilde{\theta} \rightarrow \theta_c$ with $\theta_c$ sufficiently small. In this way, the proposed ESC schemes are able to achieve practical asymptotic convergence to the extremum (see Theorems \ref{Theorem 1} and \ref{Theorem 2}).
\end{remark}  


Next, we summarize the main results in the following Theorems.
\begin{theorem} \label{Theorem 1}
Consider the closed-loop system \eqref{closed-loop system-1-2} under the Assumptions \ref{Assumption 1}, \ref{Assumption 2}, \ref{Assumption 3}, and \ref{Assumption 4}. Then, there exist class $\mathcal{K L}$ function $\beta_{a_{1}}$ with $\beta_{a_{1}} (s,0) = s$ and positive constants $\bar{a}_1, \ \Delta_1, \ k_1, \ k_2, \ \alpha_1, \ \alpha_2$ such that, for a choice of $\omega^{\prime}_H$, $\omega^{\prime}_L$, $K^{\prime}$, $\lambda_1^{\prime}$ and for each $\rho > 0$, there exist positive constants $\bar{\epsilon}_1, \ \bar{\delta}_{1}, \ \bar{\omega}_{1}, \ \bar{\gamma}_1, \ \underline{\gamma}_1$ with $\bar{\gamma}_1 \geq \underline{\gamma}_1$ such that, for all $\delta \in (0, \bar{\delta}_1), \omega \in (0, \bar{\omega}_1), \epsilon \in (0, \bar{\epsilon}_1), \gamma_1 \in (\underline{\gamma}_1, \bar{\gamma}_1)$ and for all initial conditions satisfying $a_0 \in (0, \bar{a}_1)$ and $|\left( \tilde{\bm{x}} (t_0) , \tilde{\bm{z}} (t_0)\right)| \leq \Delta_1$, the solutions of the system \eqref{closed-loop system-1-2} satisfy for all $t \geq t_0 \geq 0$
\begin{eqnarray}
|\tilde{\bm{x}} (t)| \ &\leq & \ k_1 \exp (-\alpha_1 (t - t_0)) \ |\tilde{\bm{x}} (t_0)| + \rho, \label{x_dynamics_Theorem1} \\
|\tilde{\bm{z}} (t)| \ &\leq & \ k_2 \exp (-\alpha_2 \omega \delta (t - t_0)) \ |\tilde{\bm{z}} (t_0)| + \rho, \label{z_dynamics_Theorem1} \\
|a(t)| \ &\leq & \ \beta_{a_{1}} (a_0, \ \omega \delta \epsilon (t - t_0) ), \label{a_dynamics_Theorem1} 
\end{eqnarray}
with $\tilde{\bm{x}}(t) = \bm{x}(t) - \bm{l}(\tilde{\theta} (t) + \theta^{\star} + a(t) \sin \omega t)$, $\tilde{\bm{z}} (t) = \bm{z} (t) - \bm{z}_1^p (t, a(t))$ where $\bm{z} (t) = \left(\tilde{\theta} (t), \xi (t), \tilde{\eta} (t) \right)$ and $\bm{z}_1^p (t, a(t)) =  \left(\tilde{\theta}^p (t), \xi^p (t), \tilde{\eta}^p (t) \right)$ is a unique $\left( \frac{2 \pi}{\omega} \right)$-periodic solution, characterized by $a(t)$. 
\end{theorem}
\vspace{-0.5cm}
\begin{pf}
See the Appendix for a sketch of the proof.
\end{pf}
\begin{theorem} \label{Theorem 2}
Consider the closed-loop system \eqref{closed-loop system-2-2} under the Assumptions \ref{Assumption 1}, \ref{Assumption 2}, and \ref{Assumption 3}. Then, there exist class $\mathcal{K L}$ function $\beta_{a_{2}}$ with $\beta_{a_{2}} (s,0) = s$ and positive constants $\bar{a}_2, \ \Delta_2, \ k_3, \ k_4, \ \alpha_3, \ \alpha_4$ such that, for a choice of $\omega^{\prime}_H$, $\omega^{\prime}_L$, $K^{\prime}$, $\lambda_2^{\prime}$ and for each $\rho > 0$, there exist positive constants $\bar{\epsilon}_2, \ \bar{\delta}_{2}, \ \bar{\omega}_{2}, \ \bar{\gamma}_2, \ \underline{\gamma}_2$ with $\bar{\gamma}_2 \geq \underline{\gamma}_2$ such that, for all $\delta \in (0, \bar{\delta}_2), \omega \in (0, \bar{\omega}_2), \epsilon \in (0, \bar{\epsilon}_2), \gamma_2 \in ( \underline{\gamma}_2, \bar{\gamma}_2)$ and for all initial conditions satisfying $a_0 \in (0, \bar{a}_2)$ and $|\left( \tilde{\bm{x}} (t_0) , \tilde{\bm{z}} (t_0)\right)| \leq \Delta_2$, the solutions of the system \eqref{closed-loop system-2-2} satisfy for all $t \geq t_0 \geq 0$
\begin{eqnarray}
|\tilde{\bm{x}} (t)| \ &\leq & \ k_3 \exp (-\alpha_3 (t - t_0)) \ |\tilde{\bm{x}} (t_0)| + \rho, \\
|\tilde{\bm{z}} (t)| \ &\leq & \ k_4 \exp (-\alpha_4 \omega \delta (t - t_0)) \ |\tilde{\bm{z}} (t_0)| + \rho,\\
|a(t)| \ &\leq & \ \beta_{a_{2}} (a_0, \ \omega \delta \epsilon (t - t_0) ),
\end{eqnarray}
with $\tilde{\bm{x}}(t) = \bm{x}(t) - \bm{l}(\tilde{\theta} (t) + \theta^{\star} + a(t) \sin \omega t)$, $\tilde{\bm{z}} (t) = \bm{z} (t) - \bm{z}_2^p (t, a(t))$ where $\bm{z} (t) = \left(\tilde{\theta} (t), \xi (t), \tilde{\eta} (t) \right)$ and $\bm{z}_2^p (t, a(t)) =  \left(\tilde{\theta}^p (t), \xi^p (t), \tilde{\eta}^p (t) \right)$ is a unique $\left( \frac{2 \pi}{\omega} \right)$-periodic solution, characterized by $a(t)$. 
\end{theorem}
\vspace{-0.5cm}
\begin{pf}
The proof follows directly from the proof of Theorem \ref{Theorem 1} and has been omitted. \qed
\end{pf}
\begin{remark} \label{Remark on the speeds of convergence and output convergence}
Theorems \ref{Theorem 1} and \ref{Theorem 2} state that it is possible to achieve practical asymptotic convergence to the extremum by proper tuning of the parameters in the proposed ESC schemes. Adopting the terminology introduced in Tan et al. \cite{Tan_et_al_2006} and utilized in Tan et al. \cite{Tan_et_al_2009}, the main results stated in Theorems \ref{Theorem 1} and \ref{Theorem 2} can be interpreted as follows: the solutions first converge to a small neighborhood of the set $\{ (\bm{x}, \theta) : \bm{x} - \bm{l}(\theta) = 0 \}$, then with the speed proportional to $\omega \delta$ to a neighborhood of the sets $\{ (\tilde{\theta},\xi,\tilde{\eta}): \bm{z} - \bm{z}_{i}^p = 0 \} (i=1,2)$, and finally, with the speed proportional to $\omega \delta \epsilon$ to a neighborhood of the point $ (\tilde{\theta},\xi,\tilde{\eta}, a) = (0,0,0,0)$. Thus, $ \left( \theta,\xi,\eta, a \right)$ converge to a small neighborhood of $(\theta^{\star},0,J(\theta^\star),0)$ and $y$ converges to a small neighborhood of the extremum $J(\theta^{\star})$.  
\end{remark}
\begin{remark}
It is not possible to analytically calculate $\bar{\epsilon}_i, \ \bar{a}_i, \ \bar{\delta}_{i}, \ \bar{\omega}_{i}, \ \Delta_i, \ \bar{\gamma}_i, \ \underline{\gamma}_i (i=1,2)$. However, it is possible to get conservative estimates of these quantities by carrying out experiments. These remarks are similar to the ones in Tan et al. \cite{Tan_et_al_2009}.
\end{remark}
\begin{figure}[b] 
\centering
\subfigure[The objective function]{\includegraphics[width=0.225\textwidth]{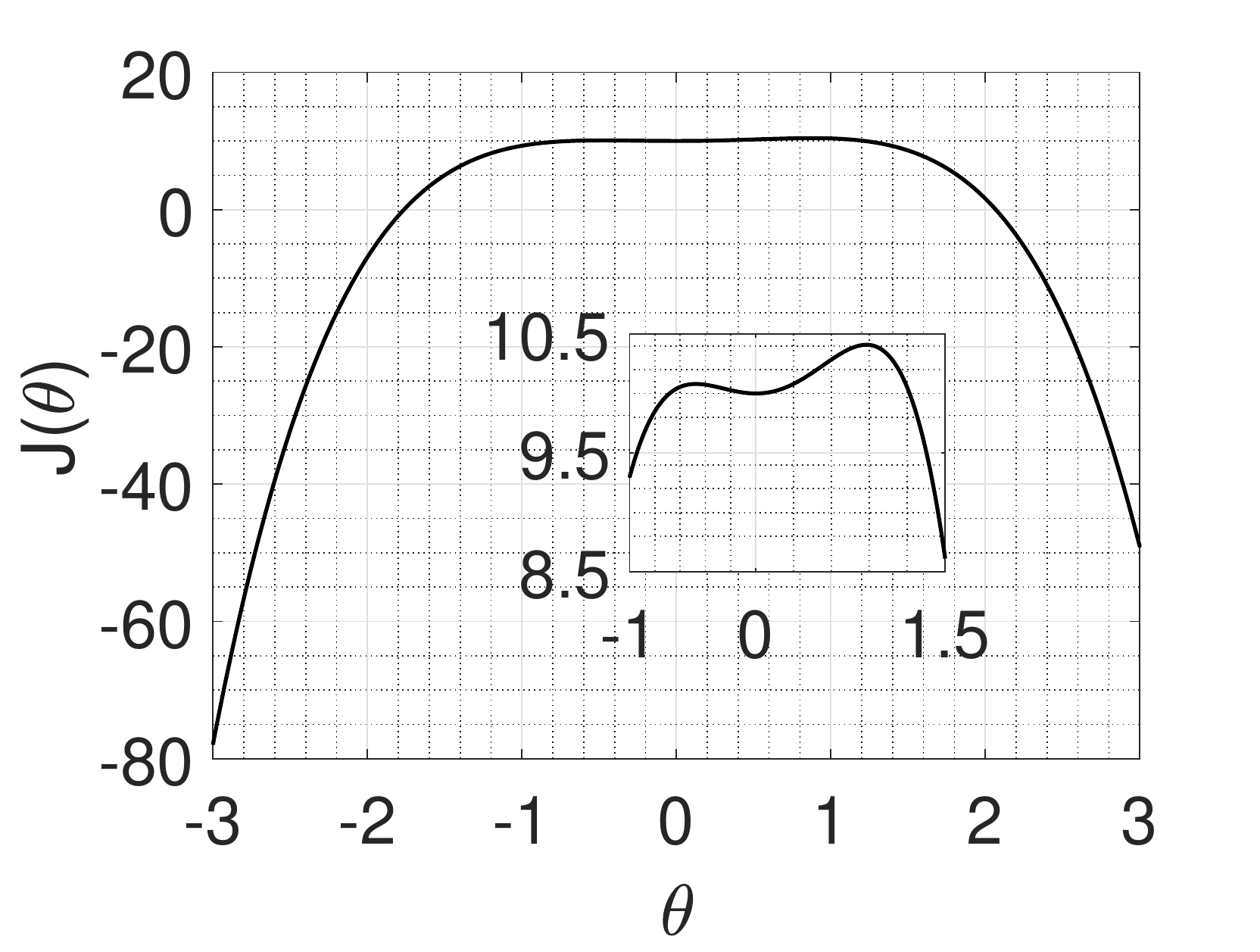}
\label{objective_function_1}}
\subfigure[The bifurcation diagram]{\includegraphics[width=0.225\textwidth]{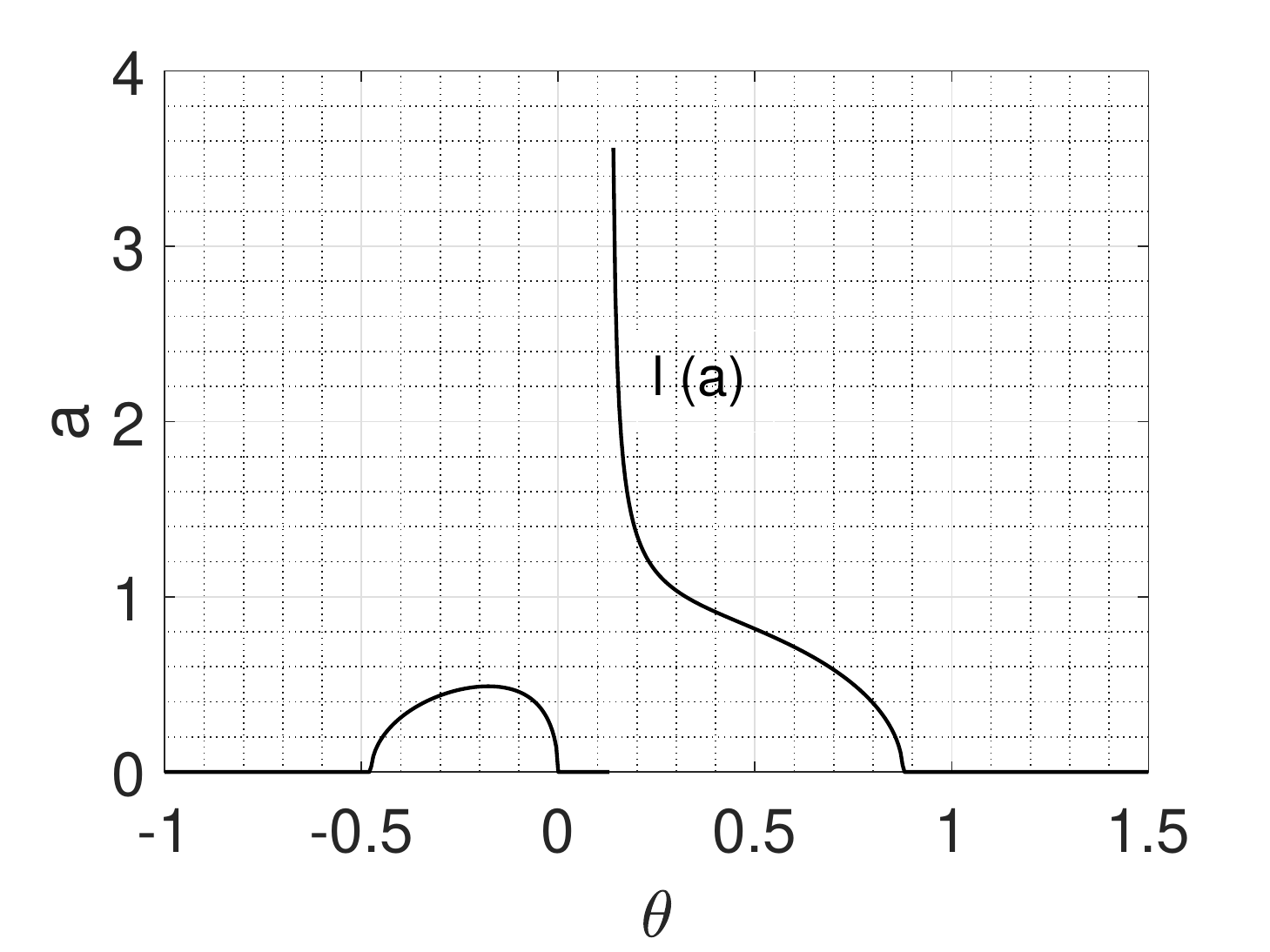}
\label{Bifurcation_diagram_1}}
\caption{The objective function in \eqref{objective function - 1 equation} and the corresponding bifurcation diagram (required for the scheme in Tan et al. \cite{Tan_et_al_2009}).}
\end{figure}
\begin{figure*}[t!]
\begin{center}
\subfigure[The output of the system]{\includegraphics[width= 0.45\textwidth]{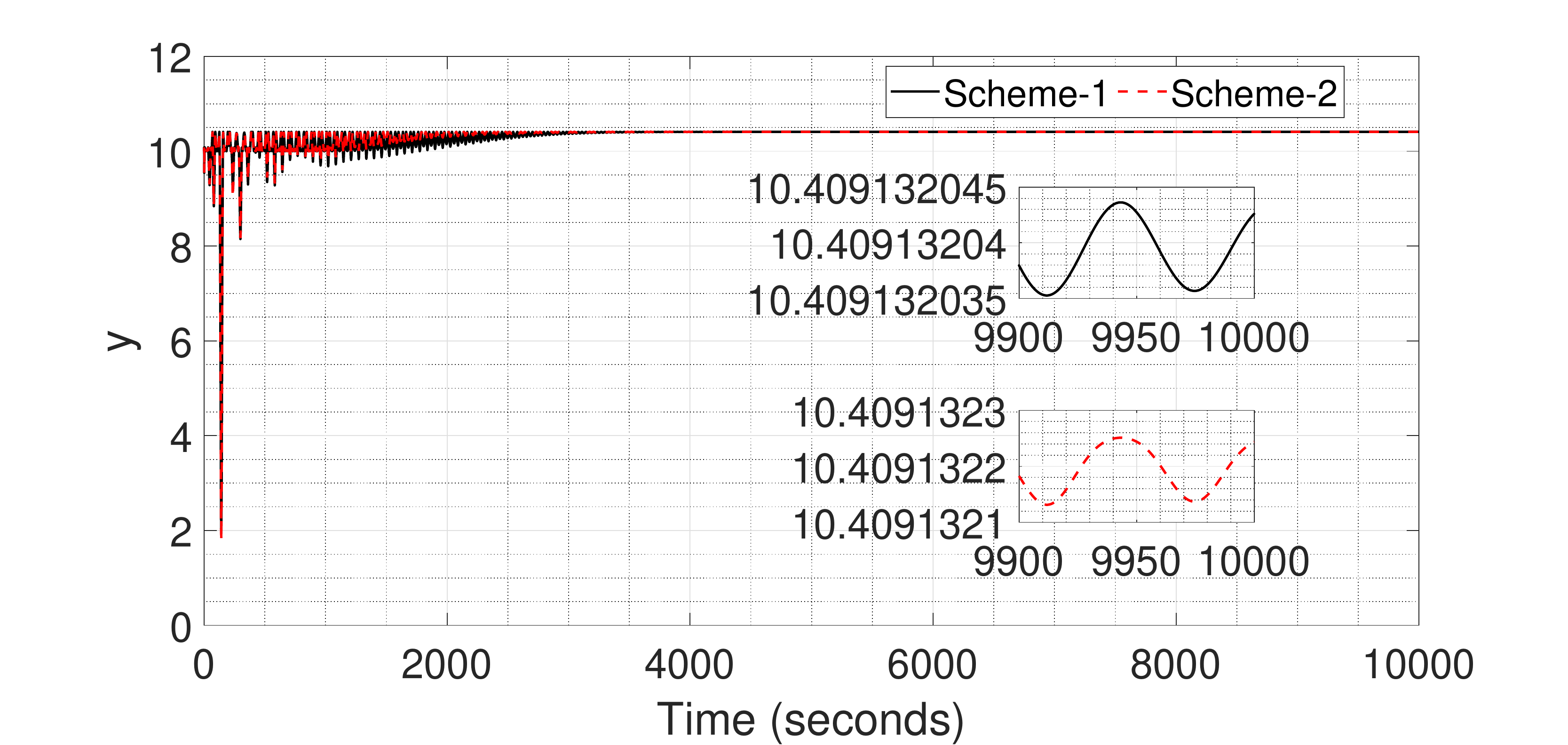} \label{Output_Proposed_schemes_obj_fn_1}} 
\subfigure[The amplitude of the excitation signal]{\includegraphics[width= 0.45\textwidth]{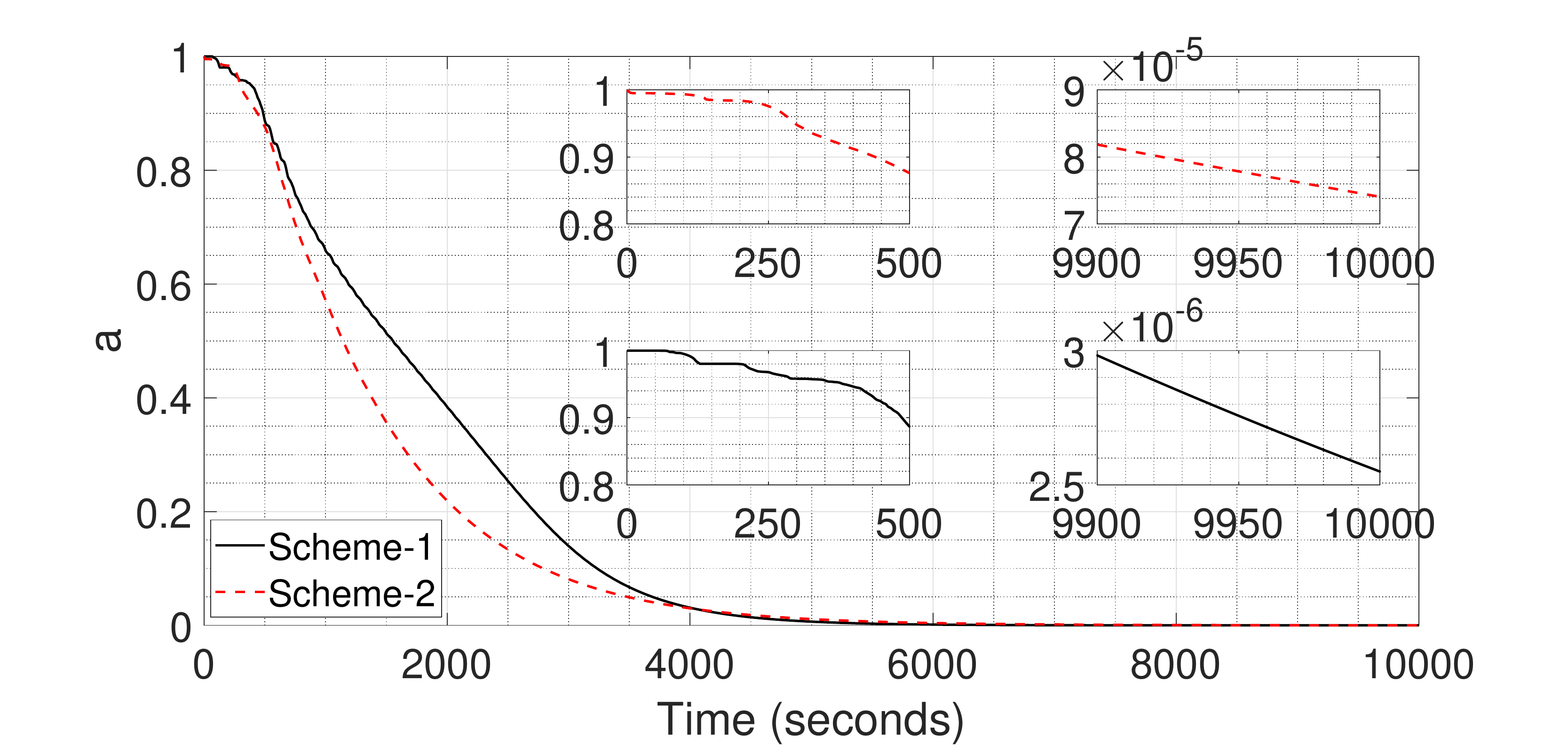} \label{Amplitude_Proposed_schemes_obj_fn_1}}
\subfigure[The estimated extremum $\hat{\theta}$]{\includegraphics[width= 0.45\textwidth]{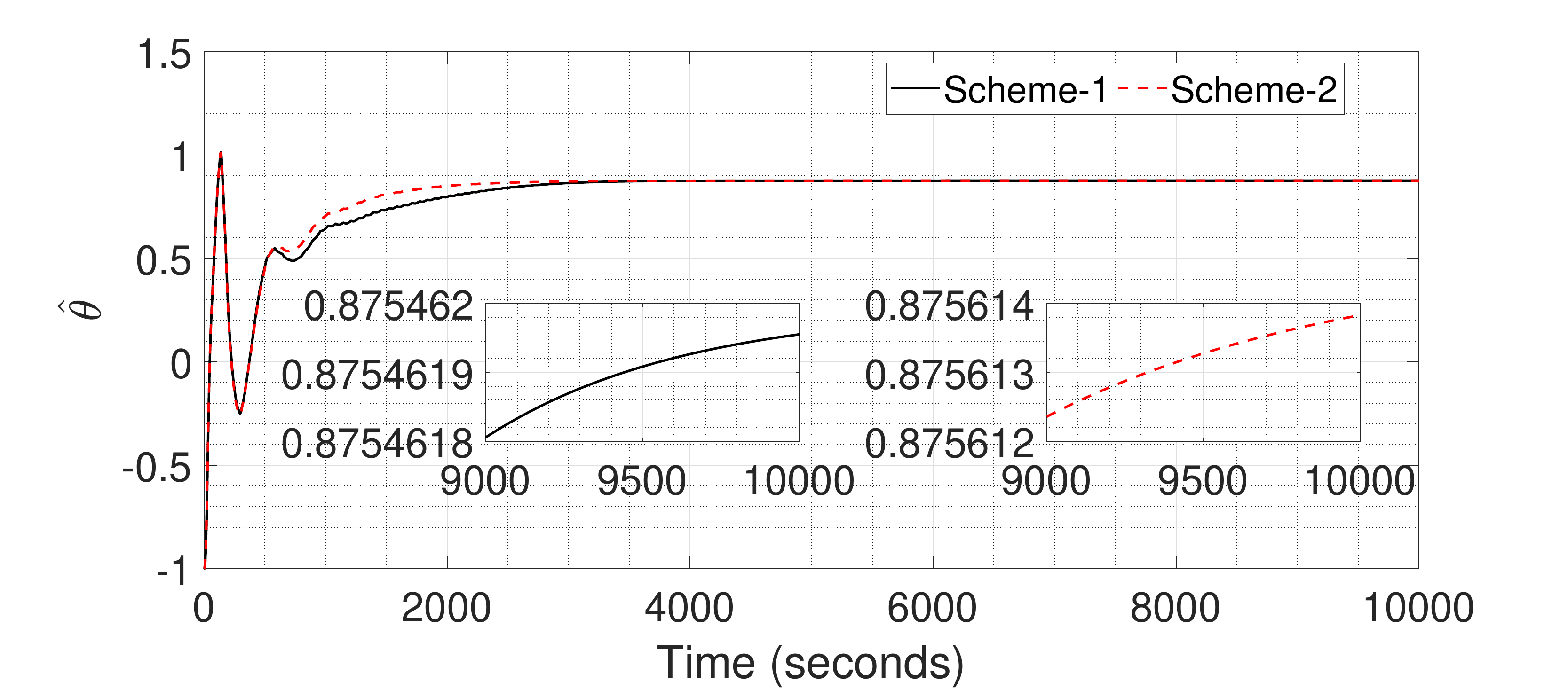} \label{Theta_hat_Proposed_schemes_obj_fn_1}} 
\subfigure[The reference $\theta$]{\includegraphics[width= 0.45\textwidth]{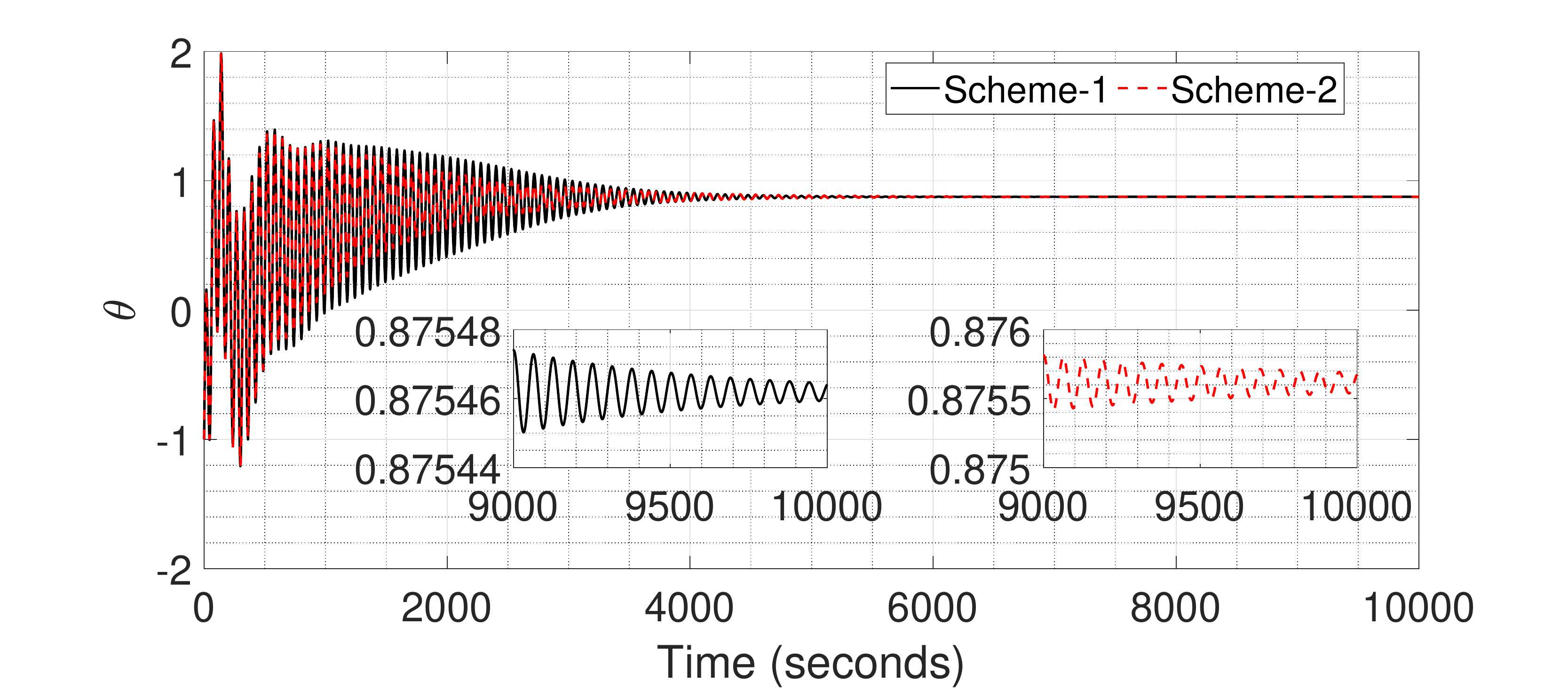} \label{Theta_Proposed_schemes_obj_fn_1}} 
\subfigure[The error in estimated gradient magnitude]{\includegraphics[width= 0.45\textwidth]{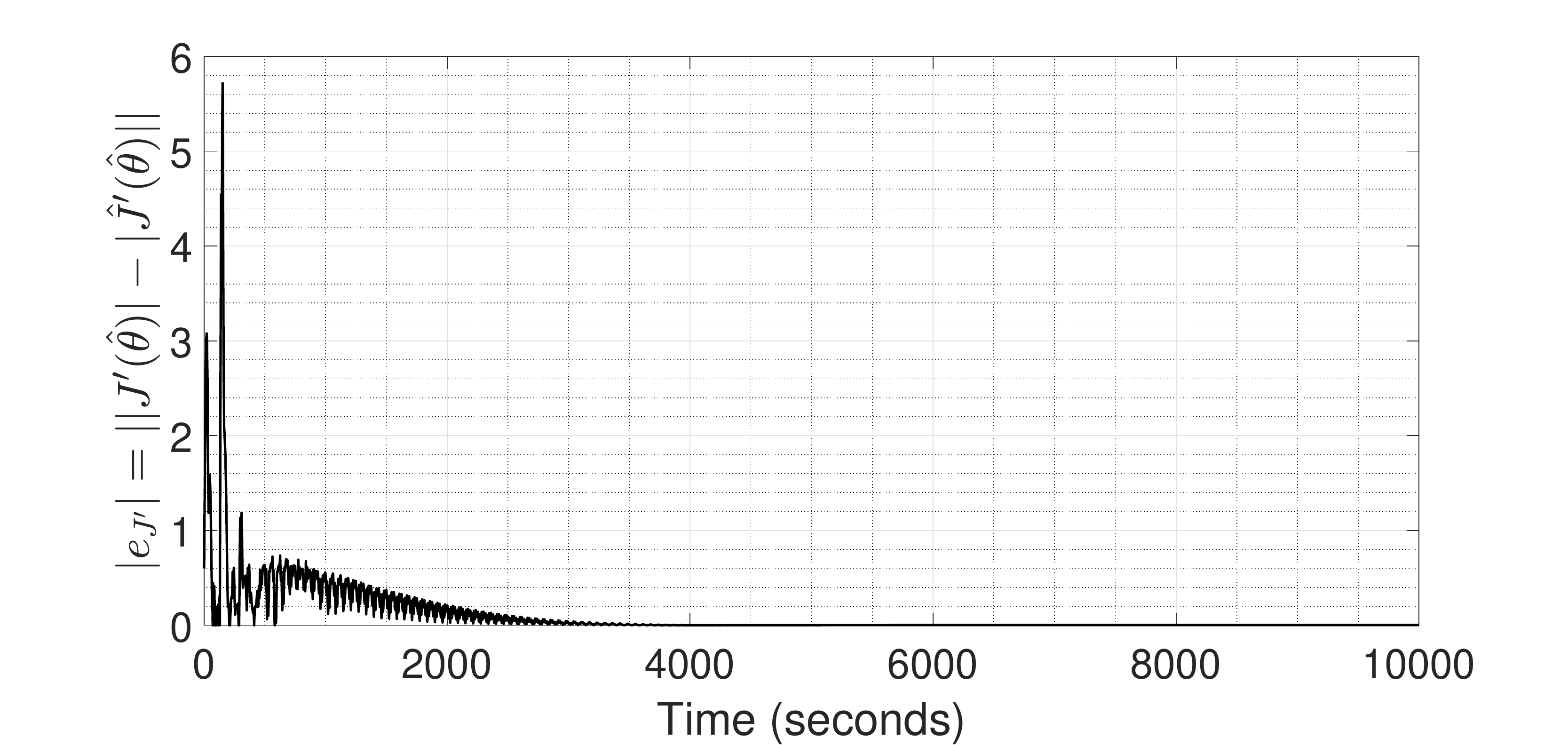} \label{Error_in_gradient_estimation_Scheme_1_obj_fn_1}}
\subfigure[The control inputs $u$]{\includegraphics[width= 0.45\textwidth]{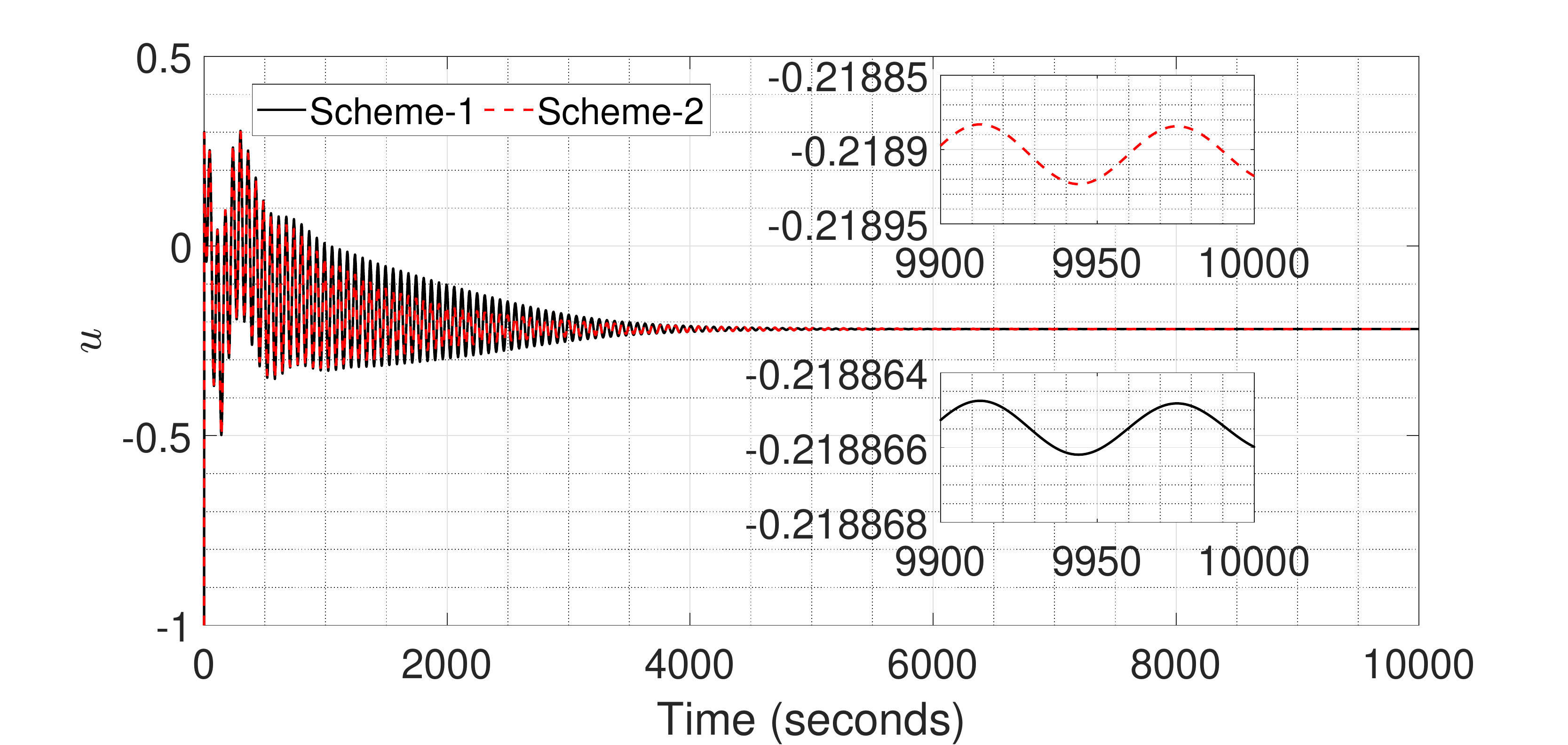} \label{Control_input_Proposed_schemes_obj_fn_1}}      
\caption{Simulation results for the proposed ESC schemes (Example-1).}  
\label{Results_proposed_schemes_obj_fn_1}                                  
\end{center}                                 
\end{figure*}
\begin{remark}  \label{Remark on the differences between the proposed schemes and the scheme in Tan et al.}
The $\hat{\theta}$ dynamics for the ESC scheme in Tan et al. \cite{Tan_et_al_2009} is given by $\dot{\hat{\theta}} = \omega \ \delta \ h(\bm{x}) \sin (\omega t)$. Clearly, as $y = h(\bm{x})$ converges to a neighborhood of $J(\theta^\star)$, $\hat{\theta}$ would (sinusoidally) oscillate with an amplitude approximately equal to $\delta J (\theta^\star)$. Thus, a sufficiently large $J (\theta^\star)$ would make $\hat{\theta}$ oscillate with a large amplitude. Although this amplitude could be reduced by selecting a small enough $\delta$, the resulting convergence speed to the extremum would reduce and additional tuning would be required. The oscillation in $\hat{\theta}$ makes $\theta = \hat{\theta} + a \sin (\omega t)$ oscillatory, even when $a \rightarrow 0$. Also, without loss of generality, we can deduce that the control input $u = \alpha(\bm{x},\theta)$, the equilibria $\bm{x} = \bm{l}(\theta)$, and $y$ would be oscillatory. On the other hand, consider the proposed schemes with $y$ in a neighborhood of $J(\theta^\star)$ and $\xi$ in a neighborhood of 0 (see Remark \ref{Remark on the speeds of convergence and output convergence}). For the proposed schemes, we have $\dot{\hat{\theta}} = k \xi$ with $k$ sufficiently small. Therefore, the oscillations in $\hat{\theta}$, $\theta$, $\bm{x} = \bm{l}(\theta)$, $y = h (\bm{x})$, and $u = \alpha(\bm{x},\theta)$ for the proposed schemes would be attenuated as the extremum is reached. Essentially, these differences in the steady-state oscillations are due to the difference in the loop structures between the proposed schemes and the scheme in Tan et al. \cite{Tan_et_al_2009} (cf. our Figs. \ref{fig1}, \ref{fig2} and Fig. 2 in Tan et al. \cite{Tan_et_al_2009}). Due to the attenuated steady-state oscillations, the proposed schemes would be more favorable, compared to the one in Tan et al. \cite{Tan_et_al_2009}, for applications where steady-state oscillations are not desirable and (or) not permitted. In addition to that, proposed ESC schemes offer the following advantages over the scheme in Tan et al. \cite{Tan_et_al_2009}: (i) $\lambda_1$ and $\lambda_2$ allow the user more control over the rate of decay in the excitation signal amplitude after the extremum is reached; (ii) the user can select the gain $k$ properly to improve the convergence speeds (Krstic \cite{Krstic_2000}, Tan et al. \cite{Tan_et_al_2006}). 
\end{remark}
   

\section{Illustrative examples}    \label{section-4}
\textbf{Example-1:} We adopt the example given in Tan et al. \cite{Tan_et_al_2009} to illustrate the performance of the proposed schemes. Consider the following SISO system
\begin{equation}
\dot{x}_1 = -x_1 + x_2, \quad \dot{x}_2 = x_2 + u, \quad y = h(\bm{x}),
\end{equation} 
where $h(\bm{x}) = -(x_1 + 3 x_2)^4 + \frac{8}{15} (x_1 + 3 x_2)^3 + \frac{5}{6} (x_1 + 3 x_2)^2 + 10$ and the control input is chosen as $u = -x_1 - 4 x_2 + \theta$. Moreover, we have the objective function given by 
\begin{equation} \label{objective function - 1 equation}
J (\theta) = - \theta^4 + \frac{8}{15} \theta^3 + \frac{5}{6} \theta^2 + 10.
\end{equation}
This function has a global maximum at $\theta^{\star} = 0.87577$, a local minimum at $\theta^{\star} = 0$, and a local maximum at $\theta^{\star} = -0.47577$ as shown in Fig. \ref{objective_function_1}. The global maximum value $J(\theta^{\star})$ is 10.409132266. 

\begin{figure*}[!t]
\begin{center} 
\subfigure[The output of the system]{\includegraphics[width= 0.4\textwidth]{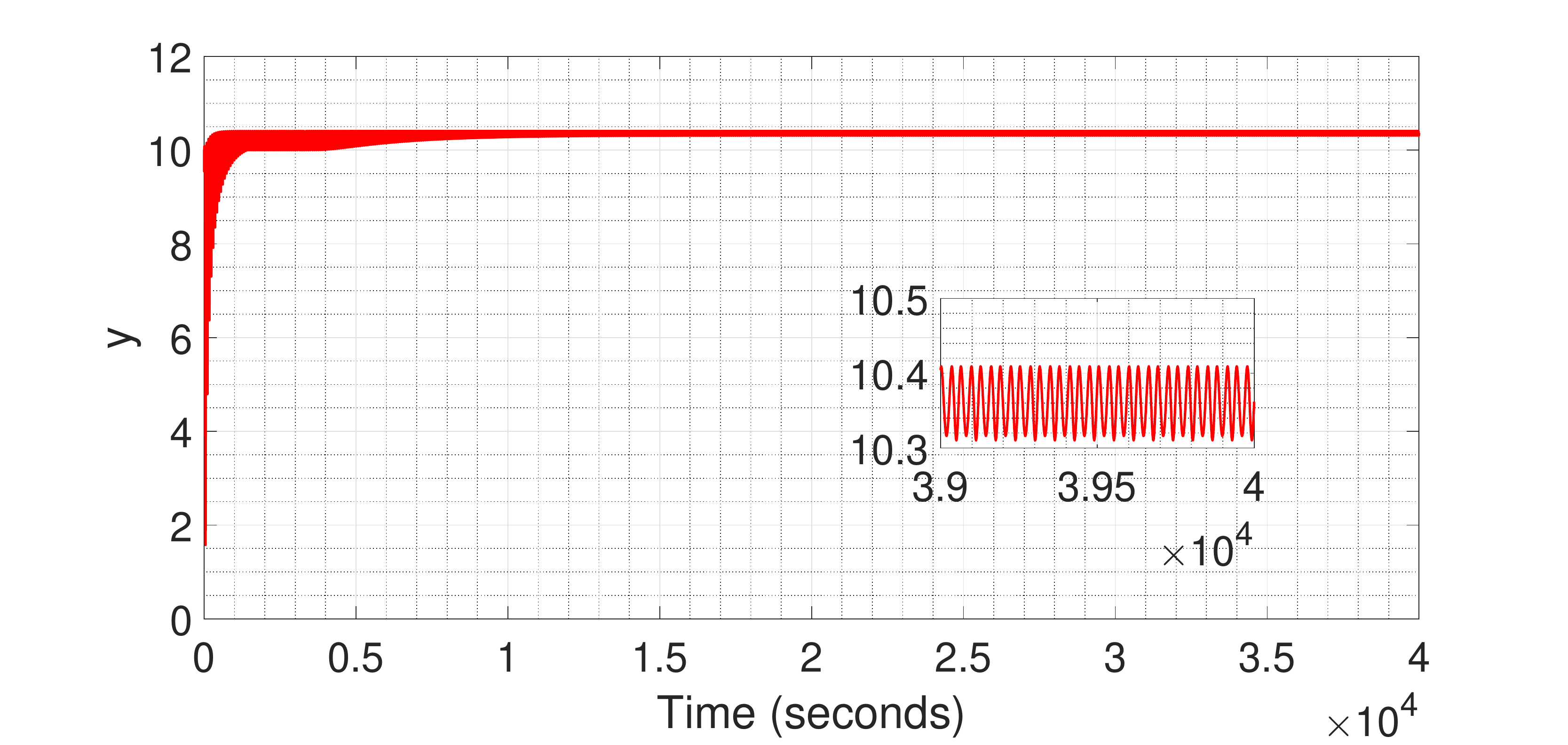} \label{Output_Tan_et_al_obj_fn_1}}
\subfigure[The amplitude of the excitation signal]{\includegraphics[width= 0.4\textwidth]{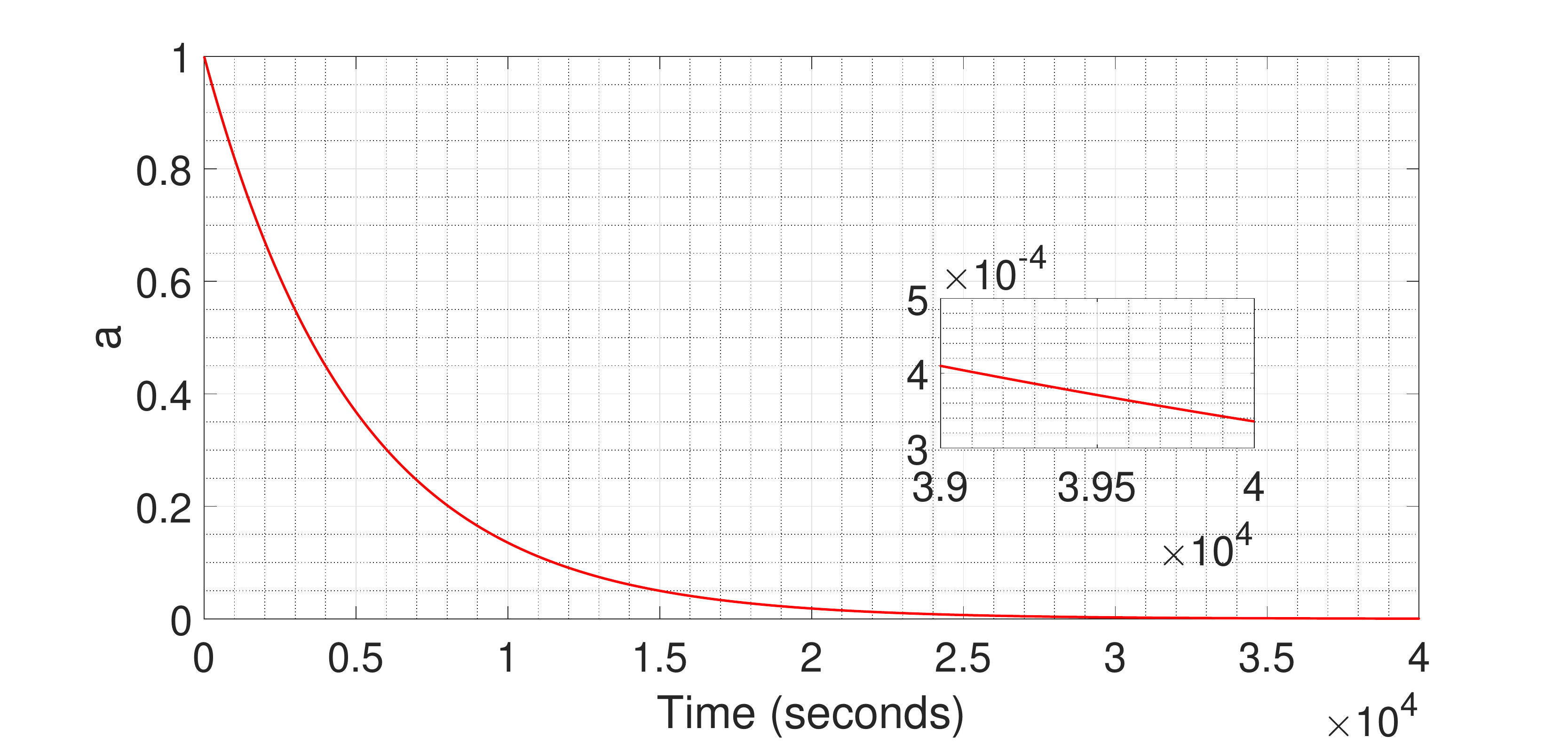} \label{Amplitude_Tan_et_al_obj_fn_1}}
\subfigure[The estimated extremum $\hat{\theta}$]{\includegraphics[width= 0.4\textwidth]{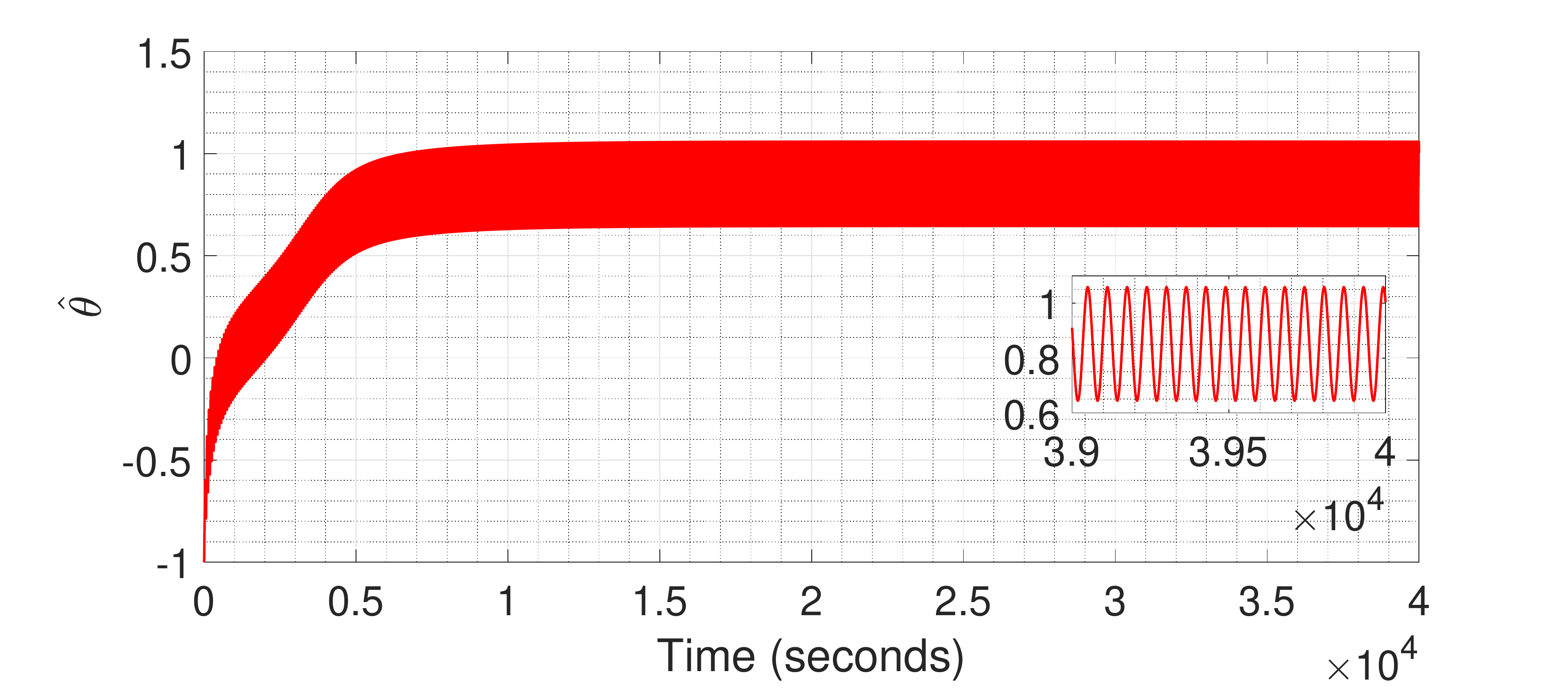} \label{Theta_hat_Tan_et_al_obj_fn_1}}  
\subfigure[The reference $\theta$]{\includegraphics[width= 0.4\textwidth]{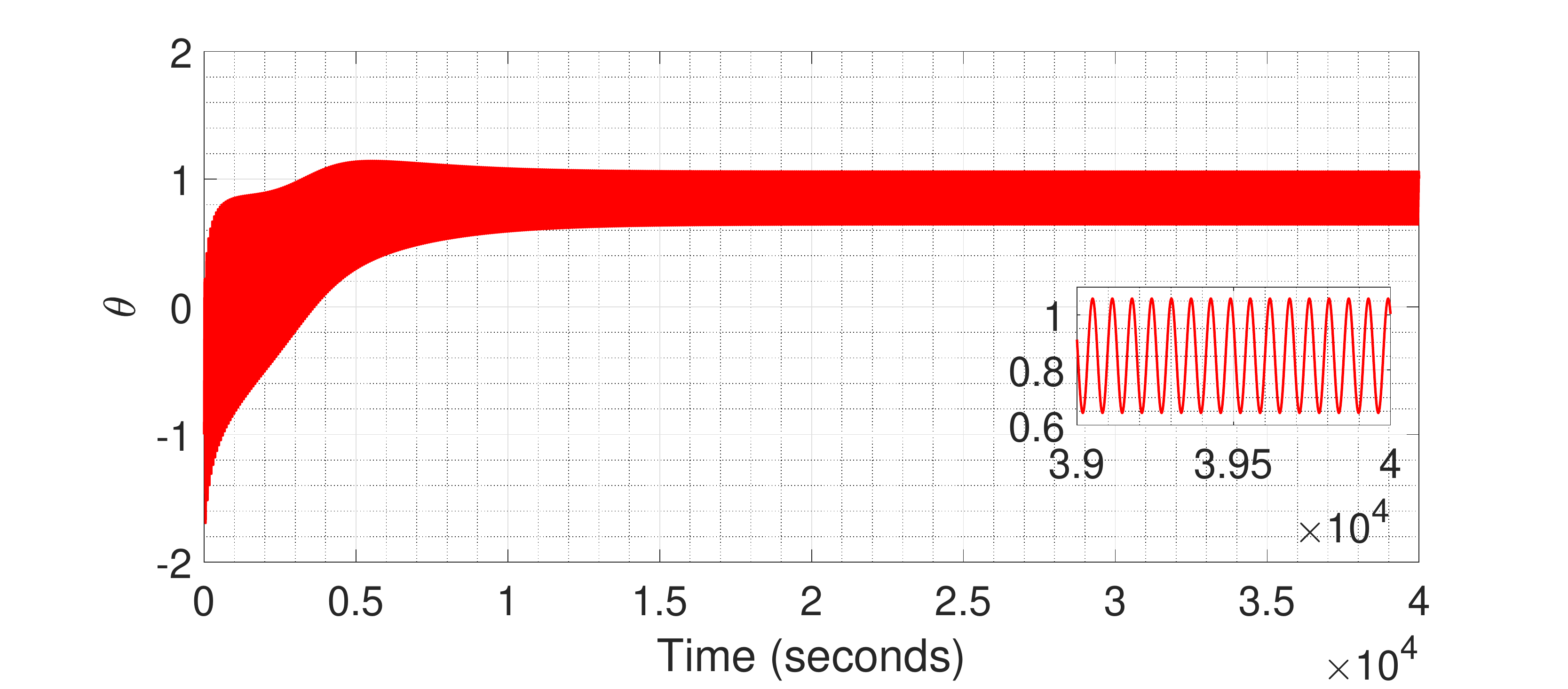} \label{Theta_Tan_et_al_obj_fn_1}}      
\caption{Simulation results for the ESC scheme in Tan et al. \cite{Tan_et_al_2009} (Example-1).}  
\label{Results_Tan_et_al_obj_fn_1}                                  
\end{center}                                 
\end{figure*} 

\begin{figure}[!hbt] 
\centering
\includegraphics[width=0.4\textwidth]{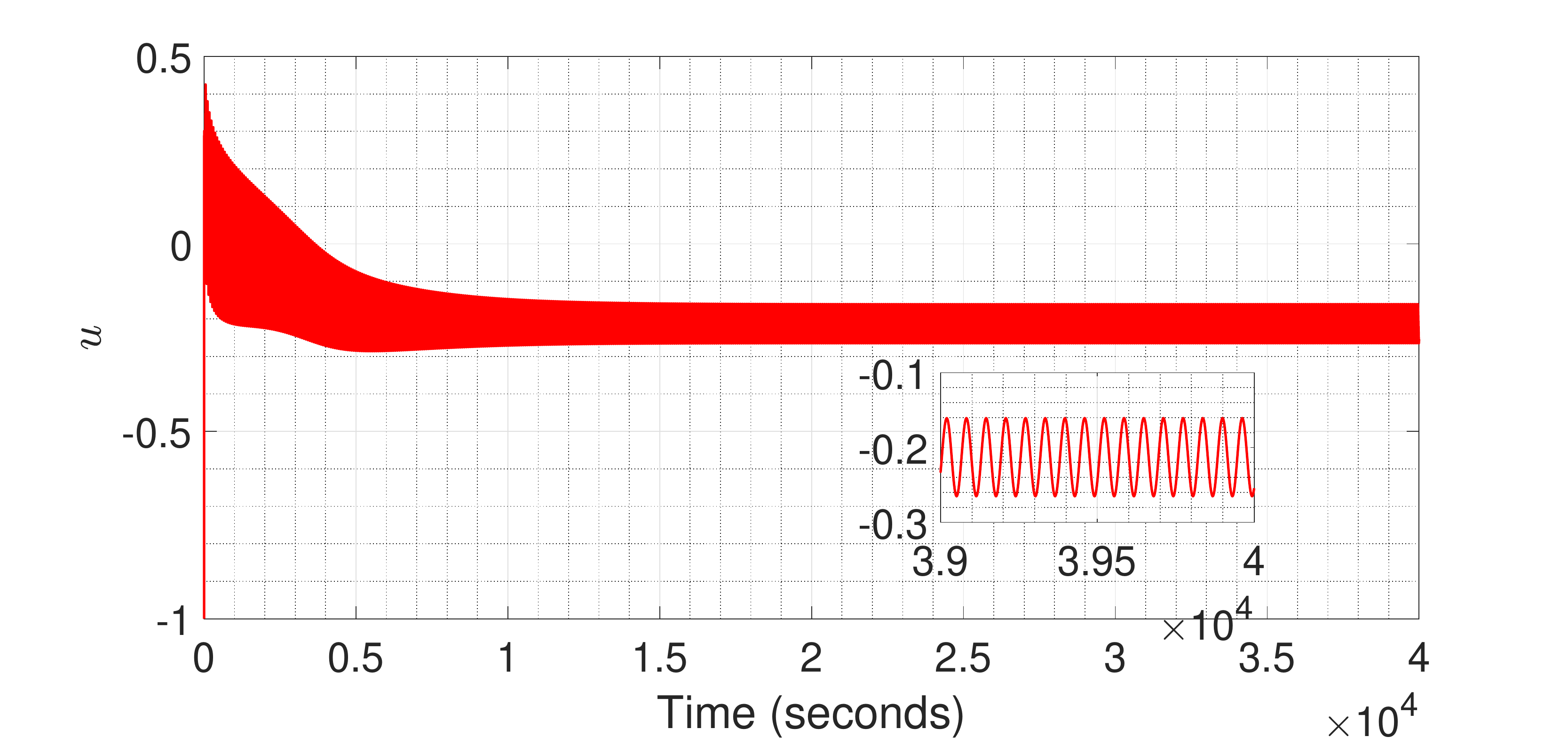}
\caption{The control input for the ESC scheme in Tan et al. \cite{Tan_et_al_2009} (Example-1).}
\label{Control_input_Tan_et_al_obj_fn_1}
\end{figure}

The simulation results for the proposed ESC schemes are shown in Figs. \ref{Output_Proposed_schemes_obj_fn_1}-\ref{Control_input_Proposed_schemes_obj_fn_1} with $t_0 = 0$, $g_1(a) = g_2(a) = a$. We choose $\hat{\theta}(t_0) = -1$ for both the schemes in order to check if the proposed ESC schemes are able to achieve global maximum despite the presence of a local maximum at $\theta = -0.47577$. The parameter values utilized for the results are as follows: (i) for both the schemes, we utilize $a_0 = 1$, $\omega^\prime_H = K^\prime = 15$, $\omega^\prime_L = 5$, $\omega = \epsilon = 0.1$, $\delta = 0.02$; (ii) for scheme-1, we take $\lambda^{\prime}_1 = 8$, $\gamma_1 = 5$, $\bm{Q} = 0.01 \bm{I}$, $r = 0.01$; (iii) for scheme-2, we take $\lambda^{\prime}_2 = 5$, $\gamma_2 = 8$. Also, for both the schemes, we choose $x_1(t_0) = x_2(t_0) = 0$. As shown in the results, both the schemes are successful in converging to the global maximum, bypassing the local extrema at $\theta = 0$ and $\theta = -0.47577$. Fig. \ref{Output_Proposed_schemes_obj_fn_1} shows that the system outputs converge to a small neighborhood of the global maximum. Also, the oscillations in the outputs are attenuated to very small amplitudes. Fig. \ref{Amplitude_Proposed_schemes_obj_fn_1} depicts the desired two phase decay of the excitation signal amplitudes (see Remark \ref{Remark on the two phase decay in the excitation signal amplitude}). In addition, $\hat{\theta}$s for both the schemes converge (approximately) within 0.035 \% of the true global maximum (Fig. \ref{Theta_hat_Proposed_schemes_obj_fn_1}) and the variations in $\hat{\theta}$s after convergence are very small, as shown in the zoomed-in plots in Fig. \ref{Theta_hat_Proposed_schemes_obj_fn_1}. $\hat{\theta}$s converge to and remain in a small neighborhood of the global maximum starting from 3000 and 2000 seconds (approximately) for scheme-1 and scheme-2, respectively (Fig. \ref{Theta_hat_Proposed_schemes_obj_fn_1}). Also, the excitation signal amplitudes are attenuated after converging to a neighborhood of the global maximum and $\theta$s converge to a neighborhood of $\hat{\theta}$s for both the schemes (Fig. \ref{Theta_Proposed_schemes_obj_fn_1}). Again, the variations in $\theta$s are attenuated as the excitation signal amplitudes decay (see the zoomed-in plots in Fig. \ref{Theta_Proposed_schemes_obj_fn_1}). The control inputs exhibit small amplitude oscillations after convergence to the extremum as well (Fig. \ref{Control_input_Proposed_schemes_obj_fn_1}).

Furthermore, Fig. \ref{Error_in_gradient_estimation_Scheme_1_obj_fn_1} shows the error associated with the estimated gradient magnitude (Kalman Filter). It is observed from Fig. \ref{Error_in_gradient_estimation_Scheme_1_obj_fn_1} that the error goes down after the initial transients. This essentially verifies the Assumption \ref{Assumption 4}. Tuning the parameters possibly would result in an improved Kalman Filter performance. But, for the proposed scheme, that is not required and selecting $\gamma_1$, $\lambda_1^\prime$ properly would suffice for convergence to the extremum, as shown in the results. 

\begin{figure}[b] 
\centering
\subfigure[The objective function]{\includegraphics[width=0.225\textwidth]{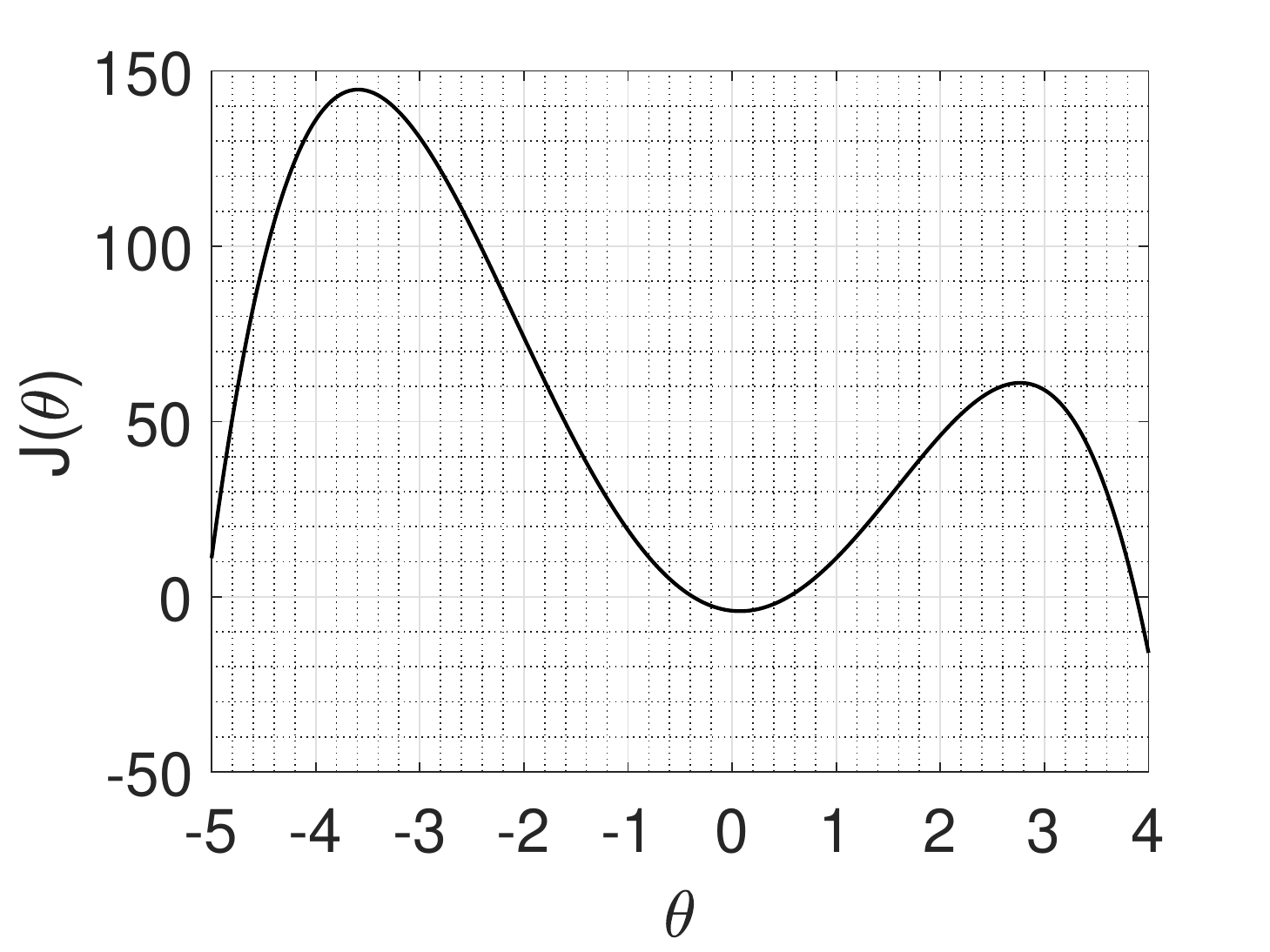}
\label{objective_function_2}}
\subfigure[The bifurcation diagram]{\includegraphics[width=0.225\textwidth]{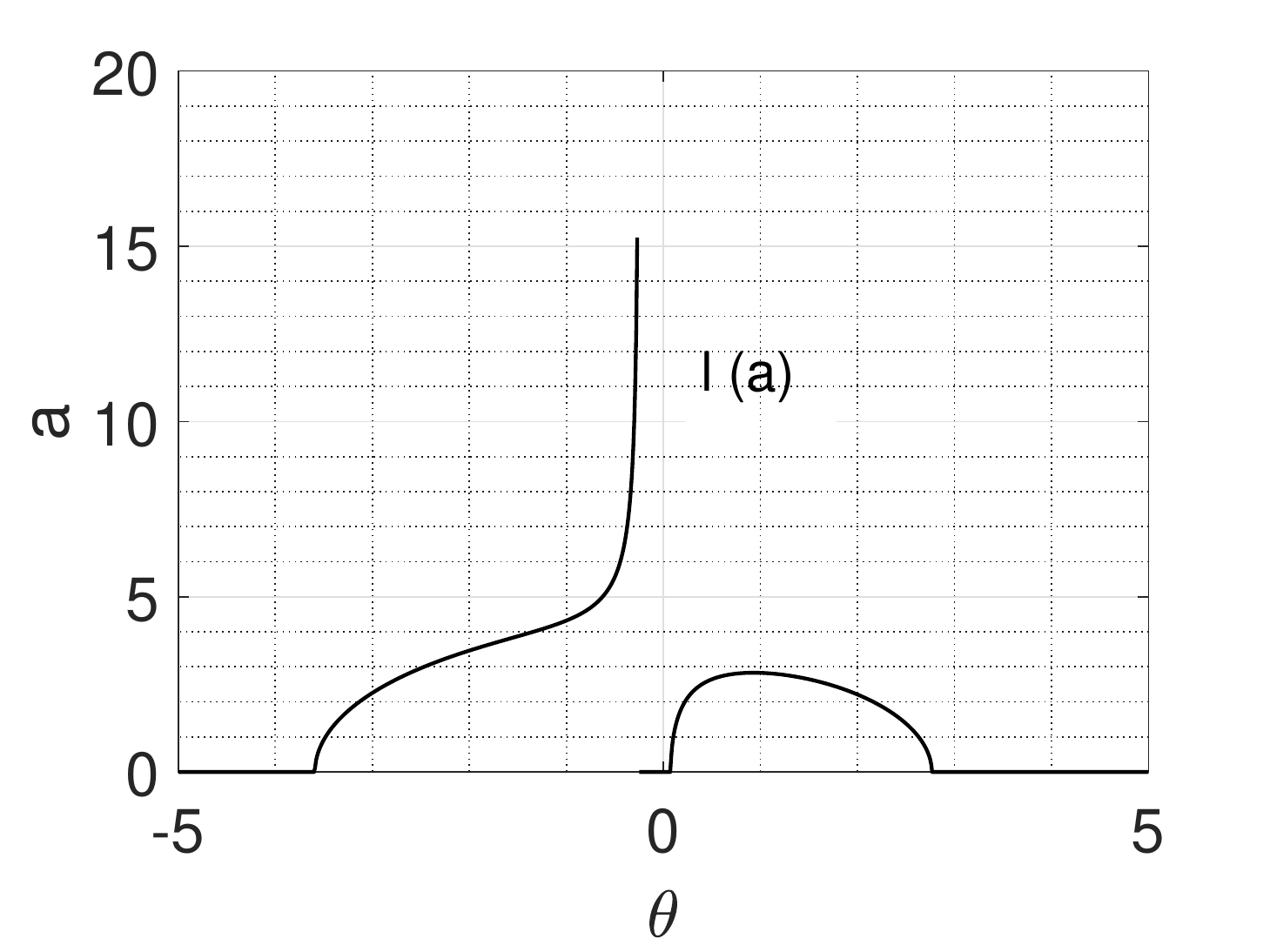}
\label{Bifurcation_diagram_2}}
\caption{The objective function in \eqref{objective function - 2 equation} and the corresponding bifurcation diagram (required for the scheme in Tan et al. \cite{Tan_et_al_2009}).}
\end{figure}

\begin{figure*}[!hbt]
\begin{center} 
\subfigure[The output of the system]{\includegraphics[width= 0.45\textwidth]{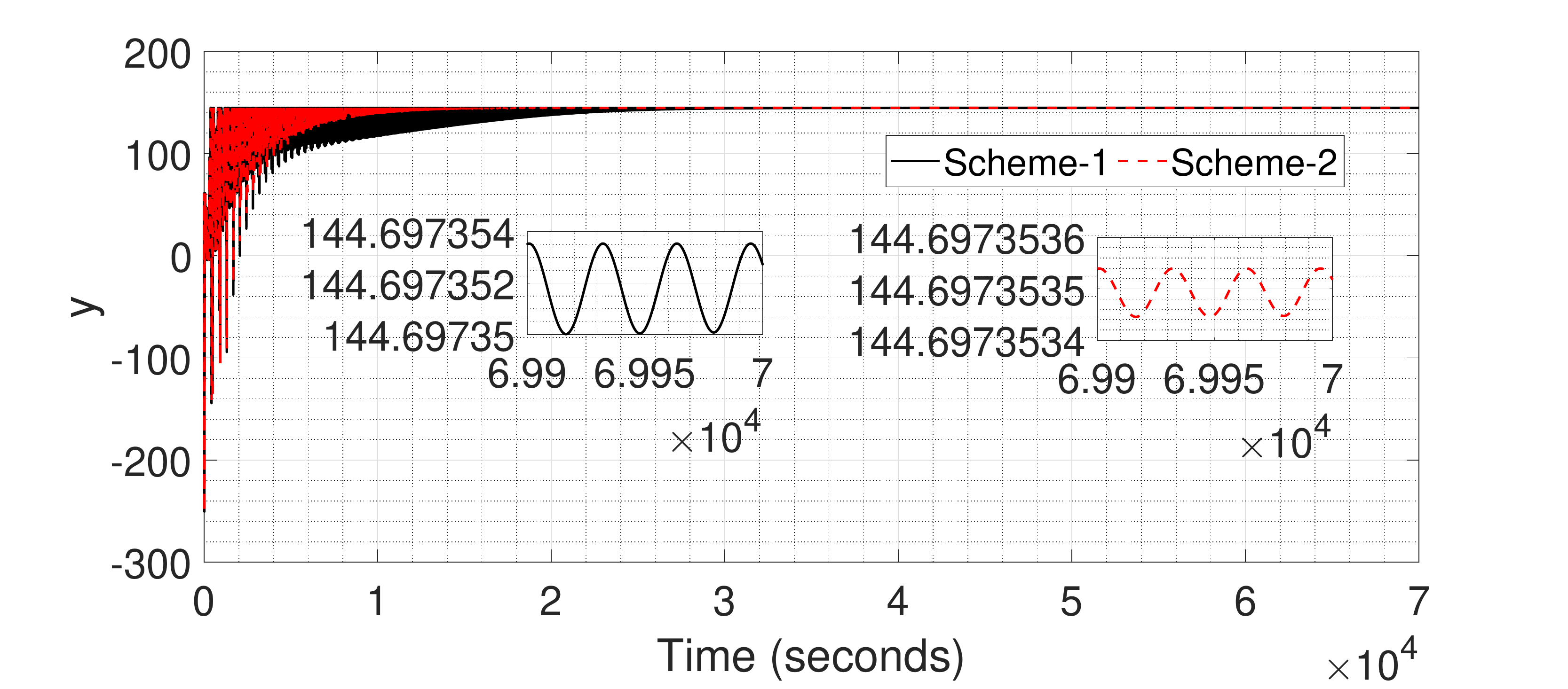} \label{Output_Proposed_schemes_obj_fn_2}}
\subfigure[The amplitude of the excitation signal]{\includegraphics[width= 0.45\textwidth]{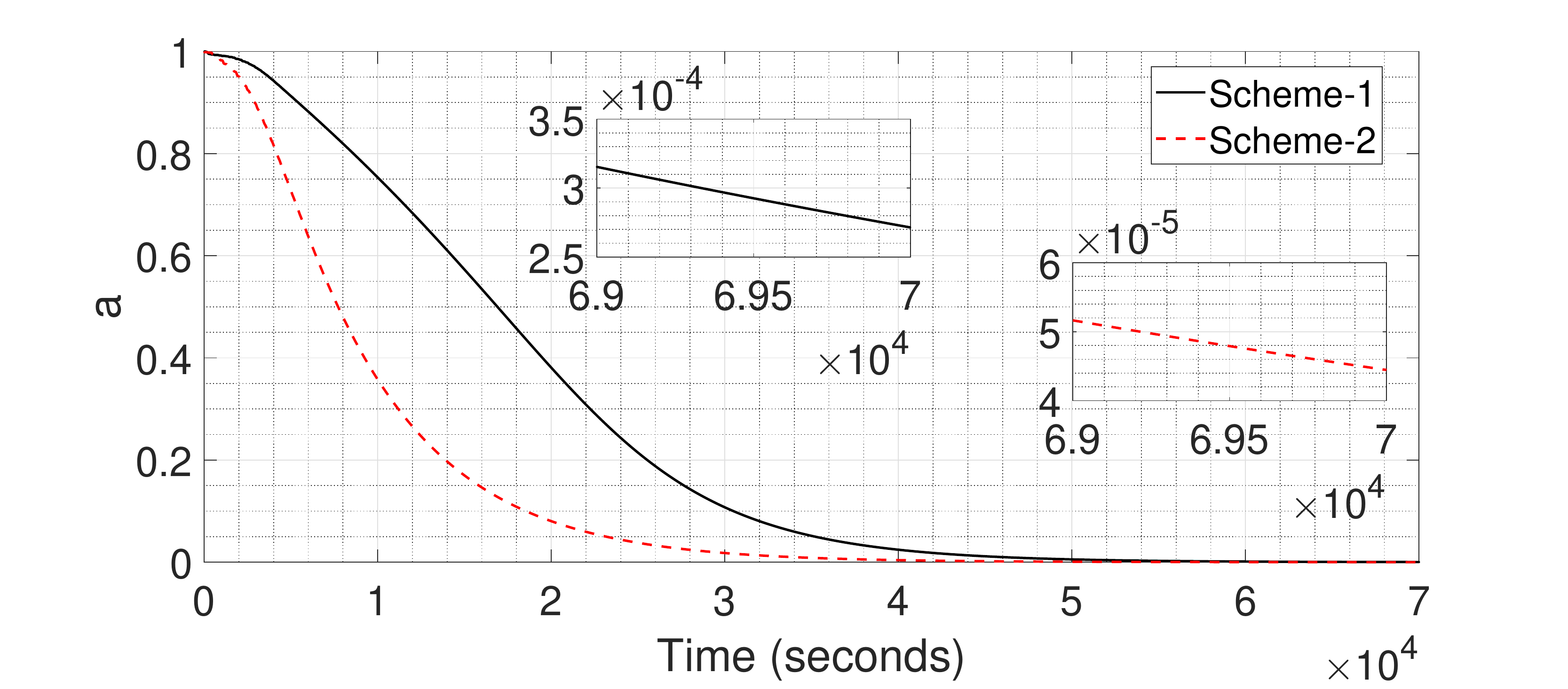} \label{Amplitude_Proposed_schemes_obj_fn_2}}
\subfigure[The estimated extremum $\hat{\theta}$]{\includegraphics[width= 0.45\textwidth]{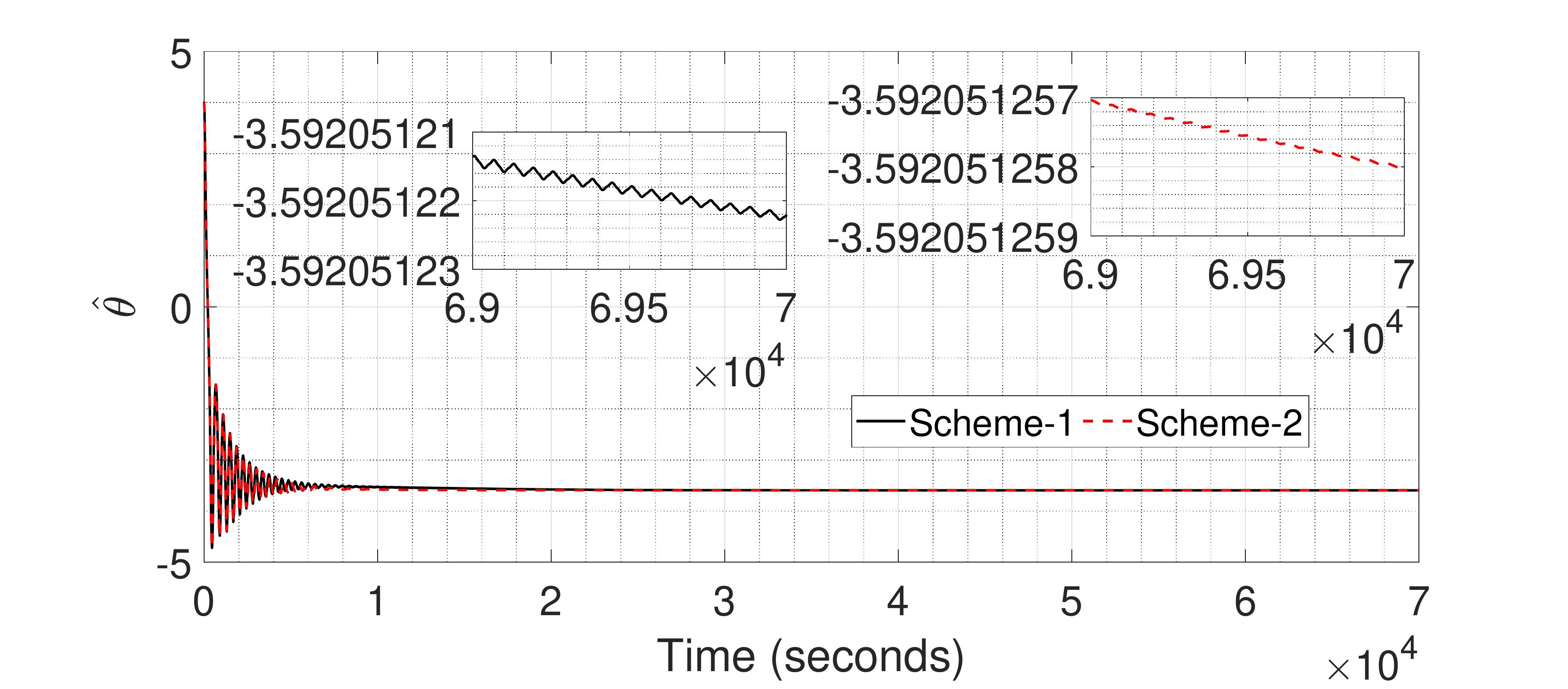} \label{Theta_hat_Proposed_schemes_obj_fn_2}}  
\subfigure[The reference $\theta$]{\includegraphics[width= 0.45\textwidth]{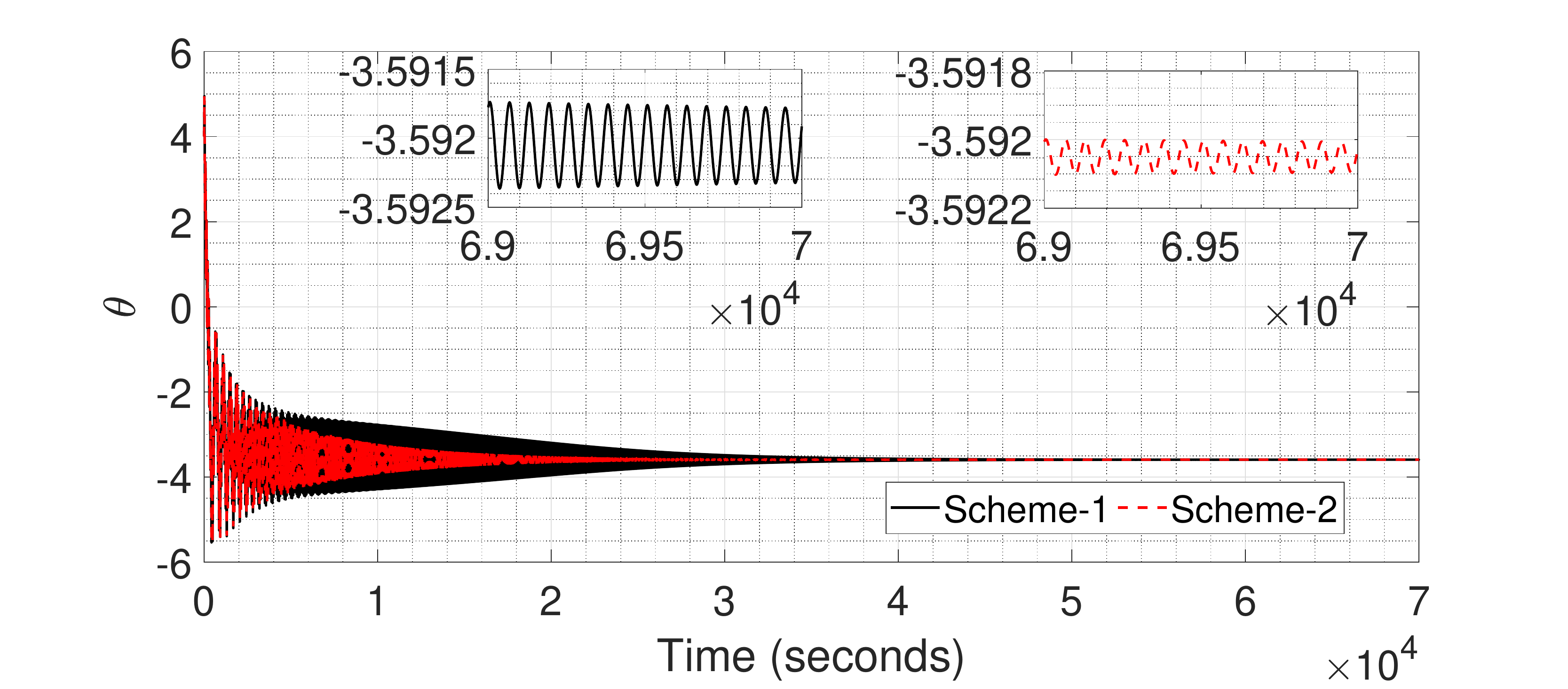} \label{Theta_Proposed_schemes_obj_fn_2}}     
\caption{Simulation results for the proposed ESC scheme (Example-2).}  
\label{Results_proposed_schemes_obj_fn_2}                                  
\end{center}                                 
\end{figure*} 

For the sake of comparison, simulation results corresponding to the scheme in Tan et al. \cite{Tan_et_al_2009} are depicted in Fig. \ref{Results_Tan_et_al_obj_fn_1} where we have utilized $\omega = \epsilon = 0.1$, $\delta = 0.02$, $g(a) = a$, $\hat{\theta}(t_0) = -1$, and $x_1(t_0) = x_2(t_0) = 0$. For these results, we have chosen $a_0 = 1$ which is sufficiently large to satisfy Assumption 4 in Tan et al. \cite{Tan_et_al_2009} (see Fig. \ref{Bifurcation_diagram_1}). It can be observed that convergence to the global extremum has been achieved and the amplitude of the excitation signal decays to a small magnitude. However, $\hat{\theta}$ and $\theta$ keep oscillating about the extremum value (Figs. \ref{Theta_hat_Tan_et_al_obj_fn_1} and \ref{Theta_Tan_et_al_obj_fn_1}). The amplitudes of these oscillations are approximately 0.2, which is equal to $\delta J(\theta^\star)$ (see Remark \ref{Remark on the differences between the proposed schemes and the scheme in Tan et al.}). As a result of the oscillations in $\theta$, the output of the system keeps oscillating about the extremum value. The control input is shown in Fig. \ref{Control_input_Tan_et_al_obj_fn_1}. Overall, the amplitudes of steady-state oscillations shown in Fig. \ref{Results_proposed_schemes_obj_fn_1} are significantly smaller compared to the results shown in Figs. \ref{Results_Tan_et_al_obj_fn_1} and \ref{Control_input_Tan_et_al_obj_fn_1}. Also, $\hat{\theta}$ converges to and remains in a neighborhood of the global maximum starting from 5000 seconds (approximately). Thus, the proposed schemes have faster convergence speeds to the extremum compared to the scheme in Tan et al. \cite{Tan_et_al_2009}. 


\textbf{Example-2:} For this example, we consider the same system as in Example-1 and choose a different objective function, given by 
\begin{equation} \label{objective function - 2 equation}
J (\theta) = - \theta^4 - \theta^3 + 20 \theta^2 - 3 \theta - 4,
\end{equation}
which has a local maximum at $\theta = 2.76658$, a local minimum at $\theta = 0.07547$, and the global maximum at $\theta = -3.59205$, as shown in Fig. \ref{objective_function_2}. The global maximum value $J (\theta^\star)$ is 144.6974. The simulation results for the proposed ESC schemes are shown in Fig. \ref{Results_proposed_schemes_obj_fn_2} with $t_0 = 0$, $g_1(a) = g_2(a) = a$. We choose $\hat{\theta}(t_0) = 4$ for both the schemes in order to check if the proposed ESC schemes are able to achieve global maximum despite the presence of a local maximum at $\theta = 2.76658$. The parameter values utilized for the results are as follows: (i) for both the schemes, we utilize $a_0 = 1$, $\omega^\prime_H = K^\prime = 6$, $\omega^\prime_L = \lambda^{\prime}_1 = \lambda^{\prime}_2 = 2$, $\omega = \epsilon = 0.1$, $\delta = 0.0075$; (ii) for scheme-1, we take $\gamma_1 = 0.1$,  $\bm{Q} = 0.01 \bm{I}$, $r = 0.01$; (iii) for scheme-2, we take $\gamma_2 = 1$. Also, for both the schemes, we choose $x_1(t_0) = x_2(t_0) = 0$. As shown in the results, both the schemes are successful in converging to the global maximum, bypassing the local extrema at $\theta = 2.76658$ and $\theta = 0.7547$. Again, the steady-state oscillations are attenuated to small amplitudes.

\begin{figure*}[!hbt]
\begin{center} 
\subfigure[The output of the system]{\includegraphics[width= 0.4\textwidth]{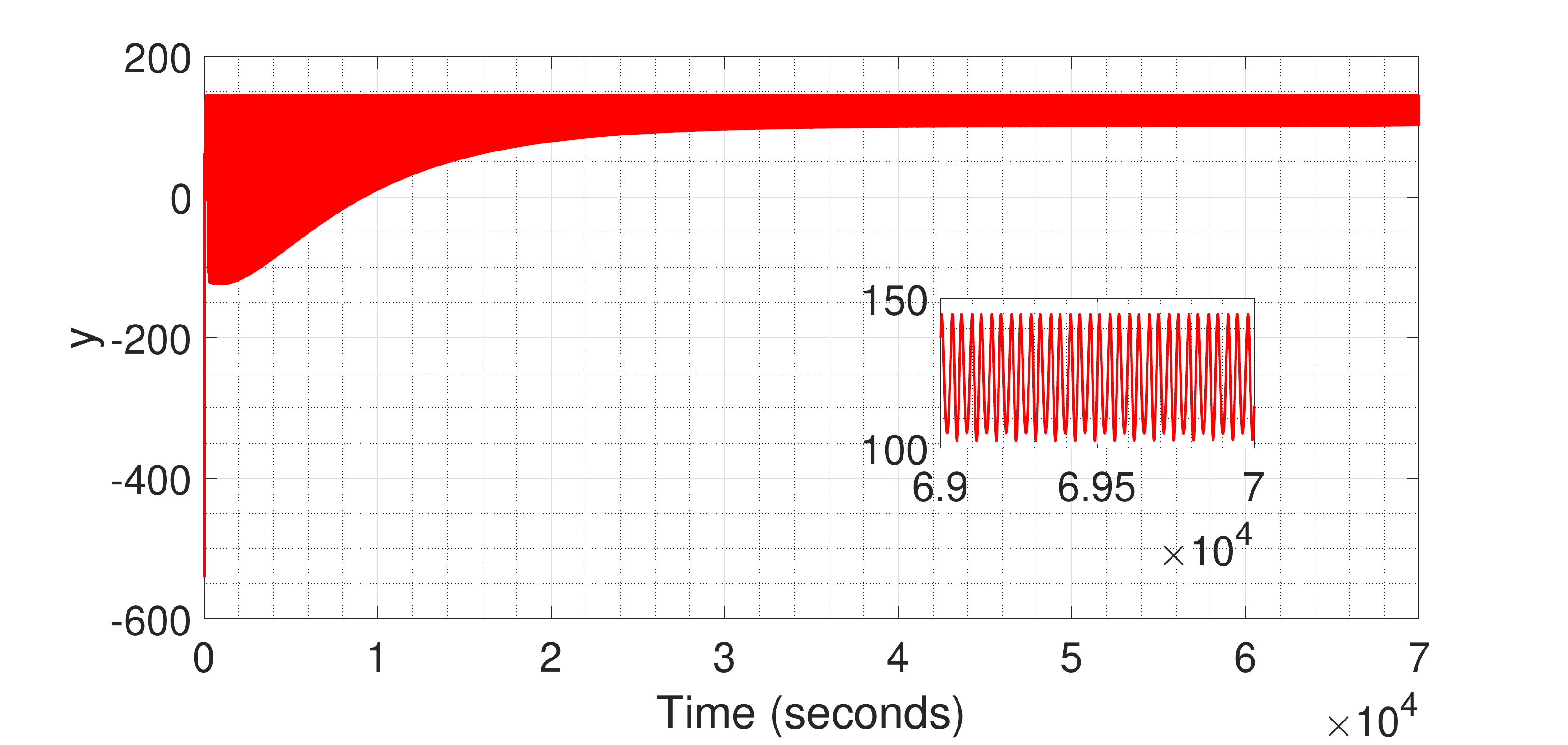} \label{Output_Tan_et_al_obj_fn_2}}
\subfigure[The amplitude of the excitation signal]{\includegraphics[width= 0.4\textwidth]{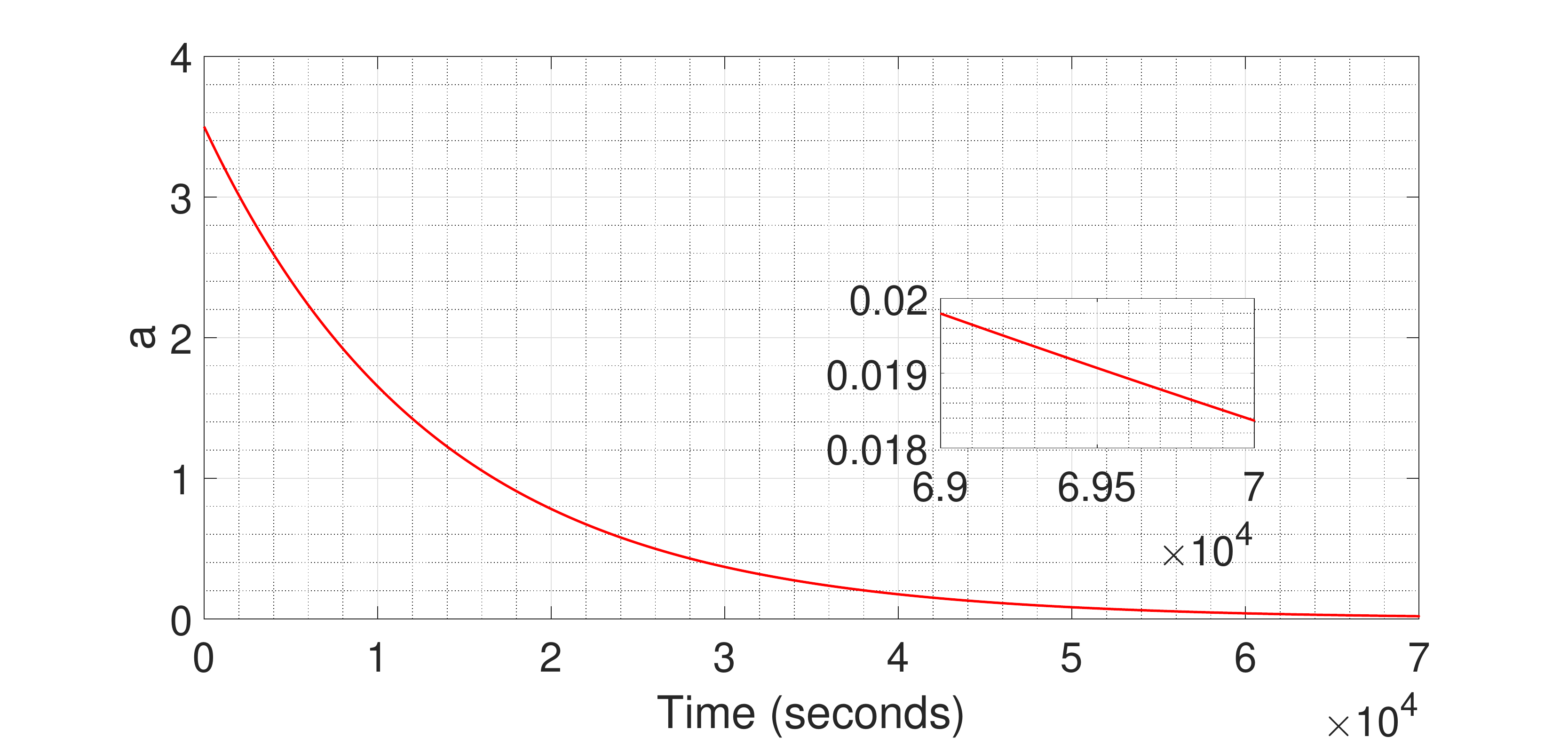} \label{Amplitude_Tan_et_al_obj_fn_2}}
\subfigure[The estimated extremum $\hat{\theta}$]{\includegraphics[width= 0.4\textwidth]{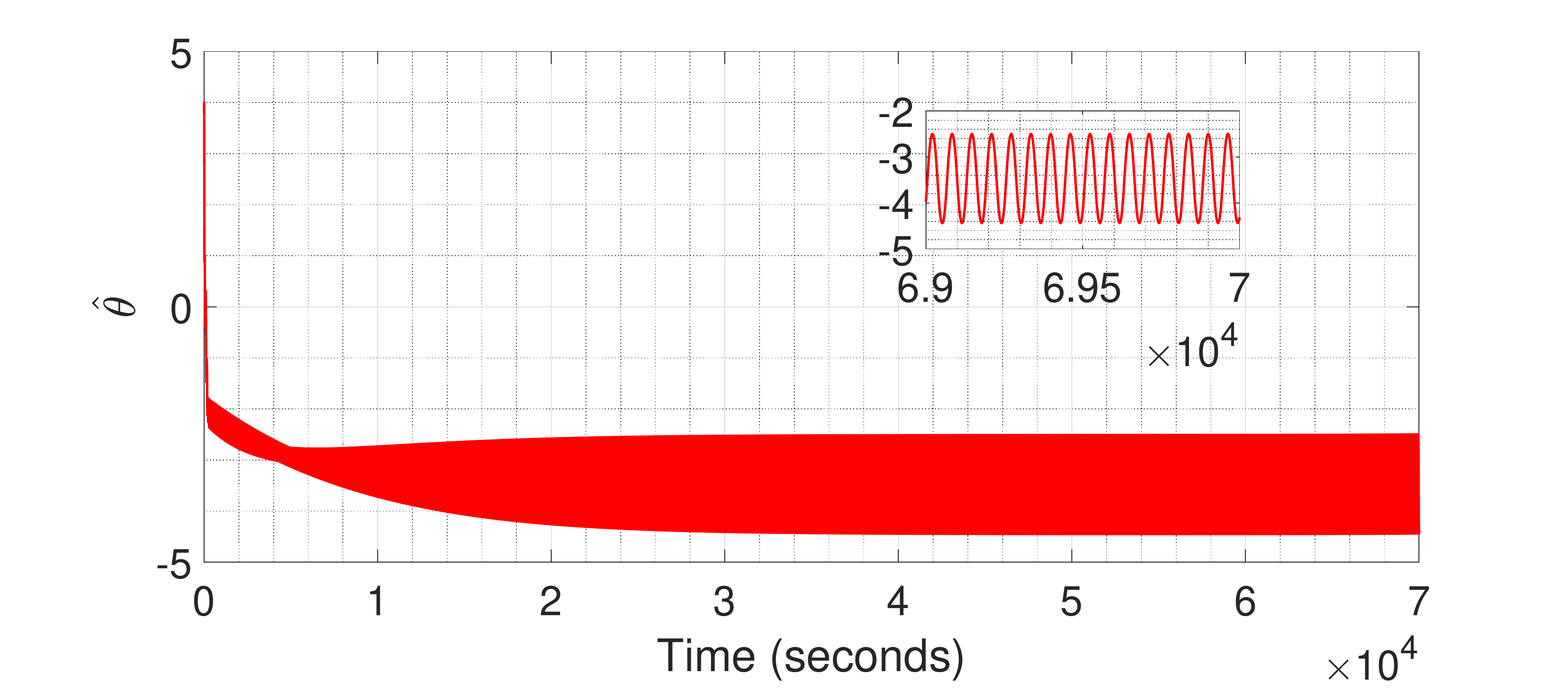} \label{Theta_hat_Tan_et_al_obj_fn_2}}  
\subfigure[The reference $\theta$]{\includegraphics[width= 0.4\textwidth]{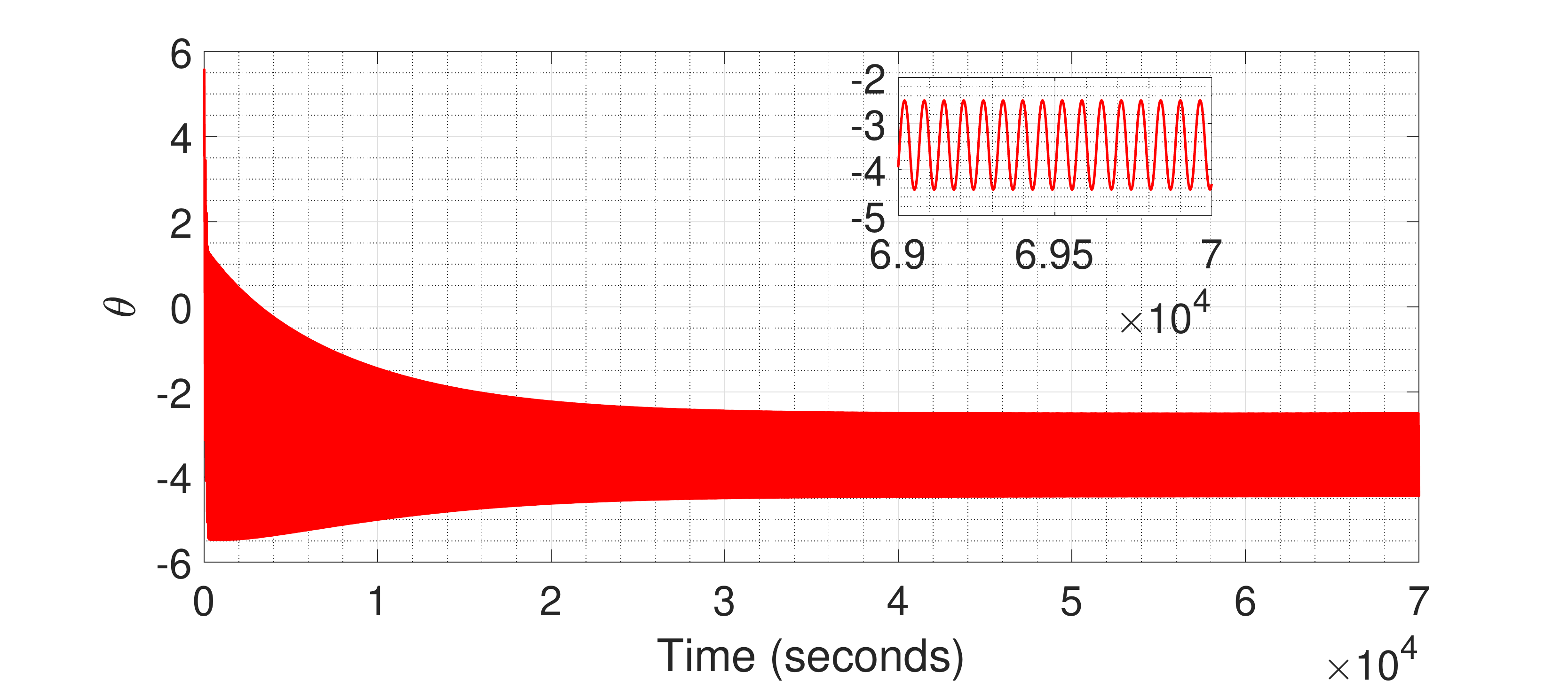} \label{Theta_Tan_et_al_obj_fn_2}}      
\caption{Simulation results for the ESC scheme in Tan et al. \cite{Tan_et_al_2009} (Example-2). }  
\label{Results_Tan_et_al_obj_fn_2}                                  
\end{center}                                 
\end{figure*} 

The results corresponding to the scheme in Tan et al. \cite{Tan_et_al_2009} are depicted in Fig. \ref{Results_Tan_et_al_obj_fn_2}. For these results, we have used $\omega = \epsilon = 0.1$, $\delta = 0.0075$, $\hat{\theta} (t_0) = 4$, and $x_1(t_0) = x_2 (t_0) = 0$ with $t_0 = 0$ seconds. Also, we choose $g (a) = a$ with $a_0 = 3.5$, which is sufficiently large to satisfy Assumption 4 in Tan et al. \cite{Tan_et_al_2009} (see Fig. \ref{Bifurcation_diagram_2}). The ESC scheme is able to converge in a neighborhood of the global maximum, as shown in these results. However, we observe that $\theta$ keeps oscillating with an amplitude of approximately 1 (which is again equal to $\delta J(\theta^\star)$ in this case), even when $a$ is reduced to small values. As a result, the output oscillates about the maximum with a large amplitude. Note that $\hat{\theta}$s converge to and remain in a small neighborhood of the global maximum starting from 6000 seconds (approximately) for both the proposed schemes (Fig. \ref{Theta_hat_Proposed_schemes_obj_fn_2}). In comparison, $\hat{\theta}$ for the scheme in Tan et al. \cite{Tan_et_al_2009} (see Fig. \ref{Theta_hat_Tan_et_al_obj_fn_2}) converges to and remains in a neighborhood of the global maximum starting from 10000 seconds (approximately). Thus, the proposed schemes have faster convergences speed to the extremum for this example as well.


\section{Summary, conclusions, and future work} \label{section-5}

In this paper, we have proposed two perturbation-based ESC schemes that are structurally similar to the classical ESC scheme. Moreover, we have proposed two novel adaptation laws for the amplitude of the excitation signal that would make the amplitude converge to zero once the extremum is reached. We have shown that it is possible for the proposed ESC schemes to achieve practical asymptotic convergence to the extremum. We have provided illustrative examples that show the effectiveness of the proposed schemes. There were two key observations from the examples: (a) the proposed schemes were able to bypass the local extremum (both local minimum and local maximum) and converge to the global maximum; (b) the steady-state oscillations in the outputs and reference inputs were attenuated to small values. The proposed methods would be applied for plume source localization problems and autonomous formation flight, both employing unmanned aerial vehicles.


\bibliographystyle{unsrt}         
\bibliography{Bibliography} 


\appendix

\section{Sketch of proof for Theorem 1}
Our proof of the result in Theorem \ref{Theorem 1} is inspired by the sketch of the proof for Theorem 2 in Nesic et al. \cite{Nesic_et_al_2012}. Therefore, we utilize the ideas given in Desoer and Shahruz \cite{Desoer_Shahruz_1986} and the generalized singular perturbation framework given in Teel et al. \cite{Teel_et_al_2003}. First, we express the closed-loop system \eqref{closed-loop system-1-2} as
\begin{equation} \label{closed-loop dynamics_proof}
\begin{split}
\dot{\bm{x}} &= \bm{f} \left( \bm{x}, \alpha(\bm{x}, \tilde{\theta} + \theta^{\star} + a \sin \omega t) \right), \\
\begin{bmatrix}
\dot{\tilde{\theta}} \\ \dot{\xi} \\ \dot{\tilde{\eta}}
\end{bmatrix}  &= \omega \begin{bmatrix}
\delta K^{\prime} \xi \\ 
- \delta \omega^{\prime}_L \xi + \delta \omega^{\prime}_L (h(\bm{x}) - \tilde{\eta} - J(\theta^{\star})) \sin \omega t \\
- \delta \omega^{\prime}_H \tilde{\eta} + \delta \omega^{\prime}_H (h(\bm{x}) - J(\theta^{\star}))
\end{bmatrix}, \\
\dot{a} &= - \omega \epsilon \left( \delta \lambda^{\prime}_1 \ g_1 (a) \ \text{exp}(-\gamma_1 |\hat{J}^{\prime}(\tilde{\theta} + \theta^{\star})|) \right).
\end{split}
\end{equation}
It is clear that the system admits three time scales with as $\omega$ and $\epsilon$ the singular perturbation parameters. The requirement for $\delta$ to be sufficiently small will be established in the subsequent analysis (cf. Proposition 1). Let us begin by putting the closed-loop system in \eqref{closed-loop dynamics_proof} into the form of (23) in Teel et al. \cite{Teel_et_al_2003}. In doing so, we get the following system.
\begin{equation} \label{closed-loop system in the form of Teel et al}
\begin{split}
\dot{\bm{x}}_{1_{s_{1}}} &= \bm{F}_{s_{1}} (\bm{x}_{s_{1}}, x_{s_{2}}, \bm{x}_f, \omega), \quad \dot{x}_{2_{s_{1}}} = \omega, \\
\dot{{x}}_{{s_{2}}} &= F_{s_{2}} (\bm{x}_{s_{1}}, x_{s_{2}}, \omega, \epsilon), \\
\dot{\bm{x}}_{f} &= \bm{F}_{f} (\bm{x}_{s_{1}}, x_{s_{2}}, \bm{x}_f), 
\end{split}
\end{equation}
where 
\begingroup
\allowdisplaybreaks
\begin{align*}
\bm{F}_{s_{1}} &= \omega \begin{bmatrix}
\delta K^{\prime} \xi \\ 
- \delta \omega^{\prime}_L \xi + \delta \omega^{\prime}_L (h(\bm{x}) - \tilde{\eta} - J(\theta^{\star})) \sin x_{2_{s_{1}}} \\
- \delta \omega^{\prime}_H \tilde{\eta} + \delta \omega^{\prime}_H (h(\bm{x}) - J(\theta^{\star}))
\end{bmatrix}, \\
\bm{x}_{s_{1}} &= \left( \tilde{\theta}, \xi, \tilde{\eta}, {x}_{2_{s_{1}}} \right) = \left( \bm{x}_{1_{s_{1}}}, {x}_{2_{s_{1}}} \right),  \\
F_{s_{2}} &= - \omega \epsilon \left( \delta \lambda^{\prime}_1 \ g_1 (a) \ \text{exp}(-\gamma_1 |\hat{J}^{\prime}(\tilde{\theta} + \theta^{\star})|) \right), x_{s_{2}} = a, \\
\bm{F}_{f} &= \bm{f} \left( \bm{x}, \alpha(\bm{x}, \tilde{\theta} + \theta^{\star} + a \sin {x}_{2_{s_{1}}}) \right), \bm{x}_{f} = \bm{x}. \\
\end{align*}
\endgroup
Note that state ${x}_{2_{s_{1}}}$ denotes time in the slower time scale $\omega t$. We define the boundary layer system as 
\begin{equation} \label{boundary layer system}
\begin{split}
\dot{\bm{x}} &= \bm{f} \left( \bm{x}, \alpha(\bm{x}, \tilde{\theta} + \theta^{\star} + a \sin {x}_{2_{s_{1}}}) \right),  \\
\dot{\bm{x}}_{s_{1}} &= 0,\\
\dot{{x}}_{s_{2}} &= 0.
\end{split}
\end{equation}
Let $\bm{x}_{\text{bl}} (t) = \left( \bm{x} (t), {\bm{x}}_{s_{1}} (t_0), x_{s_{2}} (t_0) \right)$ denote the solution to the boundary layer system \eqref{boundary layer system} starting at time $t = t_0$. Next, we define the ``average or reduced'' function-1 based on Remark 15 in Teel et al. \cite{Teel_et_al_2003}. This is achieved by ``freezing'' $\bm{x}$ at its equilibrium value $\bm{x} = \bm{l}(\tilde{\theta} + \theta^{\star} + a \sin {x}_{2_{s_{1}}})$, fixing $a$ at its initial value $a (t_0) = a_0$, and taking the following limit
\begin{equation} \label{reduced or averaged function}
\begin{split}
\bm{F}_{\text{av}_{1}} &= \lim_{\omega \rightarrow 0} \omega^{-1} \bm{F}_s (\bm{x}_s, \bm{l} (\tilde{\theta} + \theta^{\star} + a_0 \sin x_{2_{s_{1}}}) ), \\
            &= \begin{bmatrix}
							\delta K^{\prime} \xi \\
						- \delta \omega^{\prime}_L \xi + \delta \omega^{\prime}_L (\nu(\tilde{\theta} + a_0 \sin x_{2_{s}}) - \tilde{\eta}) \sin x_{2_{s_{1}}} \\
						- \delta \omega^{\prime}_H \tilde{\eta} + \delta \omega^{\prime}_H \nu(\tilde{\theta} + a_0 \sin x_{2_{s_{1}}}) \\
								1
							\end{bmatrix},
\end{split}
\end{equation}
where $\bm{F}_s = (\bm{F}_{s_{1}}, \omega)$, $\nu(\tilde{\theta} + a_0 \sin x_{2_{s_{1}}}) = J(\tilde{\theta} + \theta^{\star} + a_0 \sin x_{2_{s_{1}}}) - J(\theta^{\star})$. With our Assumption \ref{Assumption 3}, we have the following results.
\begin{eqnarray}
\nu(0) &=& 0, \\
\nu^{\prime}(0) &=& J^{\prime}(\theta^{\star}) = 0, \\
\nu^{\prime \prime}(0) &=& J^{\prime \prime}(\theta^{\star}) < 0 .
\end{eqnarray}
Thus, we define the ``average or reduced'' system-1 as
\begin{equation}  \label{medium system}
\dot{\bm{x}}_{s_1} = \omega \bm{F}_{\text{av}_{1}}, \quad \dot{x}_{s_{2}} = 0.
\end{equation}
Let $\bm{x}_{r_{1}} (t) = \left( \bm{x}_{s_1} (t), x_{s_{2}} (t_0)\right)$ denote the solution to this system starting at time $t = t_0$. We state the following result about this system.
\begin{proposition}
Consider the system \eqref{medium system} under the Assumption \ref{Assumption 3}. There exist positive constants $\bar{a}_1$ and $\bar{\delta}_1$ such that for all $a_0 \in (0, \bar{a}_1)$ and $\delta \in (0, \bar{\delta}_1)$ the solutions $\bm{x}_{1_{s_{1}}} (t) = \left( \tilde{\theta}(t), \xi(t), \tilde{\eta}(t) \right)$ exponentially converge to a unique $\left( \frac{2 \pi}{\omega} \right)$-periodic solution $\bm{x}_{1_{s_{1}}}^p (t,a_0) = \left( \tilde{\theta}^p (t), \xi^p (t), \tilde{\eta}^p (t) \right)$ satisfying
\begin{equation}
\left\lvert \begin{bmatrix}
\tilde{\theta}^p (t) + \frac{\nu^{\prime \prime \prime}(0)}{8 \nu^{\prime \prime}(0)} a_0^2 \\
\xi^p (t) \\
\tilde{\eta}^p (t) - \frac{\nu^{\prime \prime}(0)}{4} a_0^2
\end{bmatrix} \right\rvert \leq O(\delta) + O(a_0^3),
\end{equation}
for all $t \geq t_0 \geq 0$ and all $\bm{x}_{1_{s_{1}}} (t_0) \in \mathcal{B}_{\bm{x}_{1_{s_{1}}}}$ where $\mathcal{B}_{\bm{x}_{1_{s_{1}}}}$ is a closed ball centered at $\left( \tilde{\theta}(t), \xi(t), \tilde{\eta}(t) \right) = \left( - \frac{\nu^{\prime \prime \prime}(0)}{8 \nu^{\prime \prime}(0)} a_0^2 + O(a_0^3), 0, \frac{\nu^{\prime \prime}(0)}{4} a_0^2 + O(a_0^3) \right)$ and contains a neighborhood of the origin $\left( \tilde{\theta}(t), \xi(t), \tilde{\eta}(t) \right) = (0,0,0)$. 
\end{proposition}

\begin{pf}
The system \eqref{medium system} can be equivalently expressed in the slower time scale $\tau = \omega t$ as 
\begin{equation*}
\frac{d}{d \tau} \begin{bmatrix}
\tilde{\theta} \\ \xi \\ \tilde{\eta}
\end{bmatrix} = \begin{bmatrix}
							\delta K^{\prime} \xi \\
						- \delta \omega^{\prime}_L \xi + \delta \omega^{\prime}_L (\nu(\tilde{\theta} + a_0 \sin \tau) - \tilde{\eta}) \sin \tau \\
						- \delta \omega^{\prime}_H \tilde{\eta} + \delta \omega^{\prime}_H \nu(\tilde{\theta} + a_0 \sin \tau)
							\end{bmatrix}.
\end{equation*}
The rest of the proof follows from Krstic and Wang \cite[Section 4]{Krstic_Wang_2000} and noting that $t = \frac{\tau}{\omega}$. \hspace{2.5cm} \qed
\end{pf} 
\vspace{-0.5cm}
Finally, let us define the ``average or reduced'' system-2, utilizing the Remark 15 in Teel et al. \cite{Teel_et_al_2003}, as
\begin{equation} \label{reduced system -2}
\dot{a} = - \omega \epsilon \delta \lambda^{\prime}_1 \ g_1 (a) \ \text{exp}(-\gamma_1 |\hat{J}^{\prime}(\tilde{\theta}^p (t))|) ,
\end{equation}
where we have dropped $\theta^{\star}$ from the argument of $\hat{J}^{\prime} (\cdot)$ as it is a constant. Let $a (t)$ denote the solution to this system starting at $a(t_0) = a_0$. Now, let us investigate the assumptions in Teel et al. \cite{Teel_et_al_2003}. It is easy to check that the Assumptions 1 and 2 in Teel et al. \cite{Teel_et_al_2003} are satisfied. Assumption 3 in Teel et al. \cite{Teel_et_al_2003} is satisfied since solutions $\bm{x} (t)$ of \eqref{boundary layer system} locally exponentially converge to the equilibrium $\bm{l}(\tilde{\theta} (t_0) + \theta^{\star} + a_0 \sin x_{2_{s_{1}}} (t_0)) = \bm{l} (\theta (t_0))$, uniformly in $\theta (t_0)$ (due to our Assumptions \ref{Assumption 1} and \ref{Assumption 2}). With that, we choose $\omega_{f,o} (\bm{x}_\text{bl}(t))= | \bm{x} (t) - \bm{l}(\tilde{\theta} (t_0) + \theta^{\star} + a_0 \sin x_{2_{s_{1}}} (t_0))|$ and $\beta_f (\omega_{f,o}(\bm{x}_\text{bl} (t_0)), t) = k_{1} \exp (-\alpha_1 (t-t_0)) | \bm{x} (t_0) - \bm{l}(\tilde{\theta}(t_0) + \theta^{\star} + a_0 \sin x_{2_{s_{1}}} (t_0))|$ with some $k_1, \alpha_1 > 0$. Also, we take $\mathcal{H}_{f} = \mathcal{B}_{\bm{x}} \times \mathcal{B}_{\bm{x}_{1_{s_{1}}}} \times [0,\infty) \times (0,\bar{a}_1)$ where $\mathcal{B}_{\bm{x}}$ is a closed ball centered at $\bm{l}(\tilde{\theta}(t_0) + \theta^{\star} + a_0 \sin x_{2_{s_{1}}} (t_0))$ and containing a neighborhood of the point $\bm{l}(\theta^{\star})$, $\mathcal{B}_{\bm{x}_{1_{s_{1}}}}$ and $\bar{a}_1$ are as in Proposition 1. Similarly, Assumption 4 in Teel et al. \cite{Teel_et_al_2003} is satisfied with the solutions $\bm{x}_{1_{s_{1}}} (t)$ of \eqref{medium system} exponentially converging to $\bm{x}_{1_{s_{1}}}^p (t,a_0)$. Hence, we choose $\omega_{s_1,o} (\bm{x}_{r_{1}} (t)) = | \bm{x}_{1_{s_{1}}} (t) - \bm{x}_{1_{s_{1}}}^p (t,a_0) |$ and $\beta_{s_{1}} (\omega_{s_1,o}(\bm{x}_{r_{1}} (t_0)),\omega (t-t_0)) = k_2 \exp (-\alpha_2 \omega \delta (t-t_0)) | \bm{x}_{1_{s_{1}}} (t_0) - \bm{x}_{1_{s_{1}}}^p (t,a_0)|$ with some $k_2, \alpha_2 > 0$. Again, we take $\mathcal{H}_{s_{1}} =  \mathcal{B}_{\bm{x}_{1_{s_{1}}}} \times [0,\infty) \times (0,\bar{a}_1)$ where all the notations are the same as for $\mathcal{H}_{f}$. We need one additional condition for the system \eqref{reduced system -2}. From \eqref{reduced system -2} and under our Assumption \ref{Assumption 4}, it is obvious that there exists a class $\mathcal{K L}$ function $\beta_{a}$ with $\beta_{a} (s,0) = s$ such that $|a(t)| \leq \beta_{a} (|a (t_0)|,\omega \delta \epsilon (t-t_0))$ for an appropriate choice of $\gamma_1$. Therefore, we choose $\omega_{s_2,o} (a (t))= |a(t)|$, $\beta_{s_{2}} (\omega_{s_2,o}(a (t_0)),\omega \epsilon (t-t_0))= \beta_{a} (\omega_{s_2,o}(a (t_0)),\omega \delta \epsilon (t-t_0))$, and $\mathcal{H}_{s_{2}} = (0,\bar{a}_1)$. Note that all the input measuring functions in Teel et al. \cite{Teel_et_al_2003} are identically zero for our system as we do not consider any disturbances.

Introducing modifications to Assumptions 7 and 8 in Teel et al. \cite{Teel_et_al_2003} for our system are tedious and have been omitted. However, we point out that the basic properties desired by imposing the Assumptions 7 and 8 in Teel et al. \cite{Teel_et_al_2003} are staisfied. For Assumption 7 in Teel et al. \cite{Teel_et_al_2003}, we take $\mathcal{K}_f = \mathcal{H}_{f}$, $\mathcal{K}_{s_{1}} = \mathcal{H}_{s_{1}}$, and $\mathcal{K}_{s_{2}} = \mathcal{H}_{s_{2}}$. Clearly, our measuring functions are bounded on these sets. Recurrence of the sets $\mathcal{K}_{f}$, $\mathcal{K}_{s_{1}}$, and $\mathcal{K}_{s_{2}}$ is guaranteed due to the choice of these sets, the results in Krstic and Wang \cite[Theorem 5.1]{Krstic_Wang_2000}, and the monotonic decrease of $a(t)$. For Assumption 8 in Teel et al. \cite{Teel_et_al_2003}, we require the functions describing the problem to be locally Lipschitz, which is satisfied by the assumptions on the functions $\bm{f}, h, \bm{l}, g_1$. Since the variable $x_{2_{s_{1}}}$ serves the purpose of time, the Lipschitz continuity of the functions $\bm{F}_{\text{av}_{1}}$ and $\bm{F}_{s_{1}}$ in the variable $x_{2_{s_{1}}}$ should be uniform. This is satisfied since $\bm{F}_{\text{av}_{1}}$ and $\bm{F}_{s_{1}}$ are periodic in $x_{2_{s_{1}}}$ (see Remarks 26-30 in Teel et al. \cite{Teel_et_al_2003}). Also, it is easy to verify the uniform continuity requirements on the measuring functions by making proper modifications.

With regards to the solutions of \eqref{closed-loop dynamics_proof}, let us define the following change of coordinates $\tilde{\bm{x}}(t) = \bm{x}(t) - \bm{l}(\tilde{\theta} (t) + \theta^{\star} + a(t) \sin \omega t)$, $\tilde{\bm{x}}_{1_{s_{1}}} (t) = \bm{x}_{1_{s_{1}}} (t) - \bm{x}_{1_{s_{1}}}^p (t, a(t))$ where $\bm{x}_{1_{s_{1}}}^p (t, a(t))$ is the periodic solution from Proposition 1, characterized by $a(t)$. Now, by introducing suitable modifications to the main result in Teel et al. \cite{Teel_et_al_2003}, we conclude that \eqref{x_dynamics_Theorem1} and \eqref{z_dynamics_Theorem1} hold with $\bm{x}_{1_{s_{1}}} \equiv \bm{z}$ and $\bm{x}^p_{1_{s_{1}}} \equiv \bm{z}_1^p$. Further, from the dynamics of $\dot{a}$ in the system \eqref{closed-loop dynamics_proof} and under our Assumption \ref{Assumption 4}, we conclude that there exists a class $\mathcal{K L}$ function $\beta_{a_{1}}$ with $\beta_{a_{1}} (s,0) = s$ such that $a(t)$ satisfies \eqref{a_dynamics_Theorem1} for all $\gamma_1 \in (\underline{\gamma}_1, \bar{\gamma}_1)$ with some $\bar{\gamma}_1, \underline{\gamma}_1 > 0$ and $\bar{\gamma}_1 \geq \underline{\gamma}_1$. This completes the proof. \qed

\end{document}